\documentclass[prd,preprintnumbers,superscriptaddress,floatfix,twocolumn,notitlepage,nofootinbib]{revtex4-1}
\usepackage{textcase}
\usepackage{graphicx,epsfig,psfrag,amssymb,hyperref}
\usepackage{multirow}
\usepackage{color,graphicx,epsfig,psfrag,amsmath,empheq}
\usepackage{bm}
\usepackage{mathrsfs,amsfonts,, color}
\usepackage{hepunits}
\usepackage{slashed}
\usepackage[caption=false]{subfig}
\usepackage{xspace}
\usepackage{hepunits}
\usepackage{multirow}
\usepackage{enumitem}
\usepackage{ulem}
\usepackage{verbatim}
\usepackage{tabularx}
\setlist[description]{leftmargin=\parindent,labelindent=\parindent}

\usepackage{tikz}
\usetikzlibrary{trees}
\usetikzlibrary{decorations.pathmorphing}
\usetikzlibrary{decorations.markings}
\tikzset{
    photon/.style={decorate, line width=0.15mm, decoration={snake,amplitude=3pt,segment length=8pt}, draw=black},
    wino/.style={draw=redwine},    
    fermion/.style={draw=black, line width=0.2mm, postaction={decorate},
        decoration={markings,mark=at position .55 with {\arrow[draw=black,scale=2,#1]{>}}}},
    scalar/.style={draw=black, dashed,postaction={decorate},
        decoration={markings,mark=at position .55 with {\arrow[draw=black,scale=2,#1]{>}}}},
    scalarline/.style={draw=black, postaction={decorate},
        decoration={markings,mark=at position .55 with {\arrow[draw=black,scale=2,#1]{>}}}},
    gluon/.style={decorate, draw=black,
        decoration={coil,amplitude=3pt, segment length=4pt}},
    graviton/.style={decorate, draw=black,
        decoration={zigzag,amplitude=3pt, segment length=4pt}}
}
\tikzstyle{blob}=[circle,
                              minimum size=20,
                              draw=black!80,
                              fill=black!80]   
\tikzstyle{redblob}=[circle,
                              thick,
                              minimum size=0.4cm,
                              draw=red!80,
                              fill=red!60]

\newcommand{\be}{\begin{eqnarray}}
\newcommand{\ee}{\end{eqnarray}}
\def\lsim{\mathrel{\rlap{\lower4pt\hbox{\hskip 0.5 pt$\sim$}}
\raise1pt\hbox{$<$}}}

\def\sfrac#1#2{{\textstyle{#1\over #2}}}
\newcommand{\sss}{\scriptscriptstyle}
\newcommand{\lra}{\leftrightarrow}

\newcommand{\tev}{\ensuremath{\mathrm{\: Te\kern -0.1em V}}\xspace}
\newcommand{\gev}{\ensuremath{\mathrm{\: Ge\kern -0.1em V}}\xspace}
\newcommand{\mev}{\ensuremath{\mathrm{\: Me\kern -0.1em V}}\xspace}

\def\Dbar    {{\kern 0.2em\overline{\kern -0.2em \mathrm{D}}{}}\xspace}

\def\kstarbar    {{\kern 0.2em\overline{\kern -0.2em K}{}^{*0}}\xspace}

\definecolor{darkgreen}{rgb}{0,0.5,0}
\definecolor{darkblue}{rgb}{0,0,0.5}
\definecolor{newred}{rgb}{0.5,0.1,0}
\definecolor{gold}{rgb}{0.7,0.7,0}
\definecolor{newpurple}{rgb}{0.5,0,0.5}

\begin{document}

\title{Enabling Forbidden Dark Matter}

\author{James M. Cline}
\email{jcline@physics.mcgill.ca}
\affiliation{CERN, Theoretical Physics Department, Geneva, Switzerland}
\affiliation{Department of Physics, McGill University, 3600 Rue University, Montreal, Quebec, Canada H3A 2T8}

\author{Hongwan Liu}
\email{hongwan@mit.edu}
\affiliation{Center for Theoretical Physics, Massachusetts Institute of Technology, Cambridge, MA 02139, U.S.A.}

\author{Tracy R. Slatyer}
\email{tslatyer@mit.edu}
\affiliation{Center for Theoretical Physics, Massachusetts Institute of Technology, Cambridge, MA 02139, U.S.A.}

\author{Wei Xue}
\email{weixue@mit.edu}
\affiliation{Center for Theoretical Physics, Massachusetts Institute of Technology, Cambridge, MA 02139, U.S.A.}

\begin{abstract}

The thermal relic density of dark matter is 
conventionally set by 
two-body annihilations. We point out that in many simple
models, $3 \to 2$
annihilations can play an important role in determining the relic density over a broad range of model parameters.
This occurs when the
two-body annihilation is kinematically forbidden, but the $3\to 2$
process is allowed; we call this scenario {\it Not-Forbidden Dark Matter}.  We illustrate this mechanism for a vector-portal
dark matter model, showing that for a dark matter mass of
$m_\chi \sim \text{MeV - 10 GeV}$, 
$3 \to 2$ processes not only lead to 
the observed relic density, but also imply a self-interaction cross section 
that can solve the cusp/core problem. 
This can be accomplished while remaining consistent with stringent CMB constraints on 
light dark matter, and can potentially be discovered at future direct detection experiments.
\end{abstract}

\maketitle

%%%%%%%%%%%%%%%%%%%%%%%%%%%%%%%%%%%
\section{Introduction} \label{sec:intro}
%%%%%%%%%%%%%%%%%%%%%%%%%%%%%%%%%%%

\begin{figure}
\begin{center}
\resizebox{0.45\textwidth}{!}{
\begin{tikzpicture}
\node at (-0.3,1.5) {I) effective operators};
\node at (-1.1,-0.25) {DM};
\node at (2.95,-0.75) {$A'$};
\draw[line width=0.25mm ] (-1,0)--(1,0);
\draw[line width=0.25mm ] (-0.75,1)--(1,0);
\draw[line width=0.25mm ] (-0.75,-1)--(2.75,1);
\draw[photon] (1,0)--(2.75,-1);
\node[blob] at (1,0) {};
\begin{scope}[shift={(4.5,0)}]
\draw[photon] (-1,0)--(1,0);
\draw[line width=0.25mm ] (-0.75,1)--(2.75,-1);
\draw[line width=0.25mm ] (-0.75,-1)--(2.75,1);
\node[blob] at (1,0) {};
\end{scope}
\begin{scope}[shift={(0,-4.5)}]
\node at (-0.2,3.) {II) dark photon model};
\draw[photon] (1,1.5)--(1,0);
\draw[fermion] (-0.75,2.5)--(1,1.5);
\draw[fermion] (1,1.5)--(-0.75,0.5);
\draw[fermion] (-0.75,-1)--(1,0);
\draw[fermion] (1,0)--(2.75,-1);
\draw[photon] (1.75,-0.40)--(2.75,0.5);
\end{scope}
\begin{scope}[shift={(4.5,-4.5)}]
\draw[photon] (1,1.5)--(1,0);
\draw[fermion] (-0.75,2.5)--(1,1.5);
\draw[fermion] (1,1.5)--(2.75,2.5);
\draw[fermion] (-0.75,-1)--(1,0);
\draw[fermion] (1,0)--(2.75,-1);
\draw[photon] (0.25,-0.40)--(-0.75,0.5);
\end{scope}
\end{tikzpicture} 
}
\end{center}
\caption{
Schematic description of {\it Not-Forbidden Dark Matter}~(NFDM) 
paradigm. 
I) effective operators for the $3\to 2$ scattering processes; 
II) explicit model described in the text:
%ultraviolet completion of the effective theory in the
vector-portal dark matter model.  
}
\label{fig:3to2plot}
\end{figure}
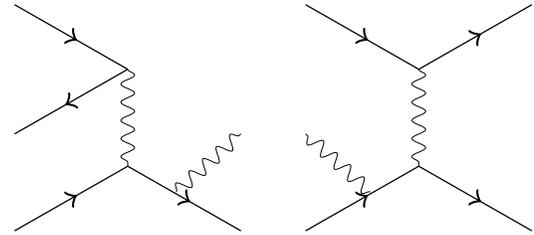

The particle physics nature of dark matter (DM) is still a mystery 
despite undeniable evidence of its gravitational interactions. The observed 
relic abundance of DM may provide a clue to its non-gravitational 
interactions, as in the classic weakly interacting massive particle 
scenario, where the freezeout of $2\rightarrow 2$ annihilation of DM particles to the standard model~(SM) particles sets the late-time abundance of DM. Many variations on the standard thermal freezeout scenario have recently been considered (e.g. \cite{Pospelov:2007mp,ArkaniHamed:2008qn, Hochberg:2014dra,Hochberg:2014kqa,Lee:2015gsa,
Hochberg:2015vrg, Bernal:2017mqb, D'Agnolo:2015pha, D'Agnolo:2015koa,Delgado:2016umt, Carlson:1992fn,Pappadopulo:2016pkp,Bernal:2015ova,Kuflik:2015isi,Bernal:2015xba,
Farina:2016llk, Dror:2016rxc, Okawa:2016wrr,
Bandyopadhyay:2011qm, D'Eramo:2010ep, Agashe:2014yua,Berger:2014sqa, Kopp:2015bfa, Berlin:2016vnh, Kopp:2016yji}); in this article, we point out that even for simple and weakly-coupled dark sectors, $3 \rightarrow 2$ annihilations -- as
illustrated in fig.~\ref{fig:3to2plot} -- can play a critical role. 

For weakly-coupled DM, $3\rightarrow 2$ processes are usually considered to be
subdominant to their $2\rightarrow
2$ counterparts at the time of freezeout,  but if
the latter are kinematically suppressed while
$3\rightarrow 2$ is unsuppressed, the situation is more
complex. This can occur when the DM couples
to a ``mediator'' particle with a mass somewhat larger than
that of the DM itself, as might arise in a
hidden sector characterized by a single scale. 

Kinematic suppression
of $2\rightarrow 2$ annihilation, leading to a novel cosmological
history during freezeout, was previously invoked
in the 
``Forbidden DM'' \cite{D'Agnolo:2015koa} and ``Impeded DM''
\cite{Kopp:2016yji} scenarios; the new feature in our study is the
presence of a kinematically allowed dark-sector $3\rightarrow 2$ annihilation channel. We
refer to this scenario as {\it Not-Forbidden Dark Matter} (NFDM).
The $3\rightarrow 2$ channel is also important in 
the Strongly Interacting Massive Particle (SIMP) scenario
\cite{Hochberg:2014dra},  but work on SIMPs has focused on strongly
coupled theories with scalar DM \cite{Hochberg:2014kqa,Choi:2016tkj}, whereas 
NFDM is a more generic mechanism: it is potentially important in any situation where $2\rightarrow 2$ annihilations within the dark sector are kinematically suppressed, and has no obvious dependence on whether the DM is fermionic or bosonic or whether the dark sector coupling is strong or weak. Hidden sector or multicomponent DM models may have regions of parameter space where NFDM is an important mechanism to consider. 

We illustrate our paradigm with a Dirac fermion DM
charged under a hidden
$U(1)$ symmetry, with dark gauge boson $A'$.  This mediator can
provide a portal to the SM by having a small coupling to
the electromagnetic current $J_{\text{EM}}^\mu$ through a kinetic mixing term $(\epsilon/2)F'_{\mu\nu} F^{\mu\nu}$. In the mass basis, the Lagrangian becomes
\begin{multline}
  \mathcal{L} \supset -\frac{1}{4} F_{\mu\nu} F^{\mu\nu} - \frac{1}{4}F_{\mu\nu}' {F'}^{\mu\nu} + \sfrac12
   { m_{A'}^2}\, {A'}_\mu {A'}^{\mu} \\
   + \bar{\chi} ( i \slashed{D} - m_\chi ) \chi   
         + e J_{\text{EM}}^\mu (A_\mu + \epsilon {A'}_{\mu})   \ . 
\label{model_def}
\end{multline}
The gauge coupling is $\alpha' = g'^2/4\pi$, and $\slashed{D} \equiv \slashed{\partial} - ig' \slashed{A'}$. It is clear in this basis that there is no tree-level coupling between $\chi$ and the SM photon. We can also consistently assume that the dark Higgs boson giving mass to
$A'$ is very heavy and can be neglected in the effective
description \cite{Kahlhoefer:2015bea}. 
Depending upon the size of the kinetic mixing parameter $\epsilon$,
there are two possible regimes of interest:

(1) $\epsilon$ is
relatively large, such that the hidden sector and the SM sector 
have the same temperature before DM freezeout, 
while $\epsilon$ is still 
small enough so that $3 \to 2$ and $2 \to 2 $ reactions involving 
only hidden sector particles dominate over annihilation of $\chi$
to SM particles;

(2) For sufficiently small $\epsilon \lesssim 10^{-8}$,
the hidden sector will have its own temperature and in the limit
 $\epsilon \to 0$, 
it becomes secluded:  
both $\chi$ and $A'$ contribute to the ultimate DM density.

In
section~\ref{Cosmology}, we  discuss the freezeout  history of
the NFDM model, 
and by solving the Boltzmann equations, we determine the dark sector
parameter values $\{m_\chi,\,m_{A'},\,\epsilon\}$ that
are consistent with the
observed relic density.   In section \ref{constraints} we incorporate
constraints from a variety of astrophysical and
laboratory searches, showing that a significant parameter region is 
allowed while realizing the NFDM mechanism.  Conclusions are given in
section \ref{sec:Summary}. In the appendix,
we present a more detailed account of how the order of freezeout of
the various reactions determines the relic abundance;
the dependence of our results on the temperature of the dark sector; how 
the constraints change with $m_{A'}/ m_\chi$,
and cross sections for the relevant scattering processes.

%%%%%%%%%%%%%%%%%%%%%%%%%%%%%%%%%%%
\section{Cosmology}
\label{Cosmology}
%%%%%%%%%%%%%%%%%%%%%%%%%%%%%%%%%%%

Previous studies of the vector-portal DM model, shown in eq.~(\ref{model_def}), 
have divided the parameter space into two broad regions: $m_\chi <
m_{A'}$ or  $m_\chi > m_{A'}$. In the latter case, the dominant
process at the epoch of thermal freezeout is $\chi \bar{\chi} \to A'
 A'$ followed by  $A' $ decays to SM particles, whereas when
$m_{\chi} < m_{A'}$, the $s$-channel annihilation  $\chi\bar\chi\to
f\bar f$ to SM particles  $f$ via off-shell $A'$ is dominant. 
This regime is ruled out for $m_\chi \sim$ MeV-GeV by
CMB constraints \cite{Adams:1998nr, Chen:2003gz, Padmanabhan:2005es, Ade:2015xua, Slatyer:2015jla}.

In the present work, however, we are interested in the intermediate region 
$m_{\chi } \lesssim m_{A'}$, where it is possible for  the $3\to 2$
scatterings $\chi \chi \bar{\chi} \to \chi A' $ or $\chi \bar\chi
 A' \to \chi  \bar{\chi}$ to have an important effect on the dark
matter abundance.  The system is governed by the coupled
Boltzmann equations for the $\chi$ and $A'$ densities. For $m_\chi \lesssim m_{A'}$, the relevant terms in these equations are
\begin{multline}
   \frac{ d n_\chi} { d t } + 3 H n_\chi =  - \frac{1}{4} \langle \sigma v^2
\rangle_{\chi\chi\bar{\chi}\atop  \to \chi A'}  
 \left(  n_\chi^3 - n_{\sss \chi,0}^2 n_\chi  \frac{n_{\sss A'}} {n_{\sss A', 0}}\right) \\
         + \langle \sigma v \rangle_{A'A'\atop \to \bar{\chi} \chi} 
        \left(  n_{\sss A'}^2  -  n_{\sss A', 0}^2 \frac{n_\chi^2}{n_{\sss \chi,0}^2}     \right), \\
\label{eq:boltz1}
\end{multline}

\begin{multline}
  \frac{ d n_{\sss A'}} { d t } + 3 H n_{\sss A'} =
        \frac{1}{8} \langle \sigma v^2
\rangle_{\chi\chi\bar{\chi}\atop  \to \chi A'  }
          \left(  n_\chi^3 - n_{\sss \chi,0}^2 n_\chi \frac{n_{\sss A'}} {n_{\sss A',0}}\right)  \\
         - \langle \sigma v \rangle_{A'A'\atop \to \bar{\chi} \chi} 
        \left(  n_{\sss A'}^2  -  n_{\sss A',0}^2 \frac{n_\chi^2}{n_{\sss \chi,0}^2}     \right) 
   - \Gamma_{A'\to f\bar f}\left( n_{\sss A'} -n_{\sss A',0}\right),
\label{eq:boltz2}
\end{multline}

where $n_\chi\,(n_{\chi,0})$ denotes the (equilibrium) density of 
$\chi+\bar\chi$,
and similarly $n_{A'}\,(n_{A',0})$ for the dark photon. Throughout this paper, we have assumed zero chemical potential for $\chi$ and $\overline{\chi}$, and take the densities of $\chi$ and $\overline{\chi}$ to be equal.
The $1/4$ in the first term of eq. (\ref{eq:boltz1}) is the symmetry factor for Dirac DM, taking into account the two identical particles in the initial state and the fact that each annihilation process removes a $\chi \overline{\chi}$ pair. The conjugate process $\chi \overline{\chi} \overline{\chi} \to \overline{\chi} A'$ is also accounted for in this factor. The relative numerical factors between the two equations are consistent with the way each process changes the number density of $\chi$ and $A'$; for example, the factor of $1/4$ and $1/8$ in the first terms of eq.~(\ref{eq:boltz1}) and (\ref{eq:boltz2}) respectively are consistent with the fact that the $3 \to 2$ process has a net effect of removing a $\chi \overline{\chi}$ pair and producing a single $A'$. A detailed discussion of the derivation of the Boltzmann equation for $3 \to 2$ processes can be found in \cite{Bernal:2015bla}. 
 
Other $3 \to 2$ processes such as $\chi \overline{\chi} A' \to \chi \overline{\chi}$, $3A' \to \chi \overline{\chi}$ etc. are important only in the case of 
$m_{A'}/m_\chi < 1$ and $\epsilon=0$ in Sec.~\ref{sec:shd}. The complete Boltzmann equations containing all of these processes are shown in eq. (\ref{eq:fullboltz1}) and (\ref{eq:fullboltz2}) in Appendix \ref{app:boltz}. All numerical results in this paper across the full range of $m_{A'}/m_\chi$ considered were obtained using the complete equations.  Expressions for 
the cross sections are given in Appendix \ref{app:xsects}. 

We will focus on the two regimes where (1) 
the hidden
sector and the SM remain in thermal equilibrium, requiring
values of the kinetic mixing $\epsilon \gtrsim 10^{-7}$ (but still
small enough to avoid dominance of the  $\chi \bar{\chi}  \to
e^+e^-$ process); (2) the hidden  sector is
secluded from the SM, $\epsilon \to 0$.

%%%%%%%%%%%%%%%%%%%%%%%%%%%%%%%%%%%
\subsection{Kinetic equilibrium with the SM}
%%%%%%%%%%%%%%%%%%%%%%%%%%%%%%%%%%%

\begin{figure*}[t!]
\subfloat[]{
\label{fig:23decay}
\includegraphics[scale=0.41]{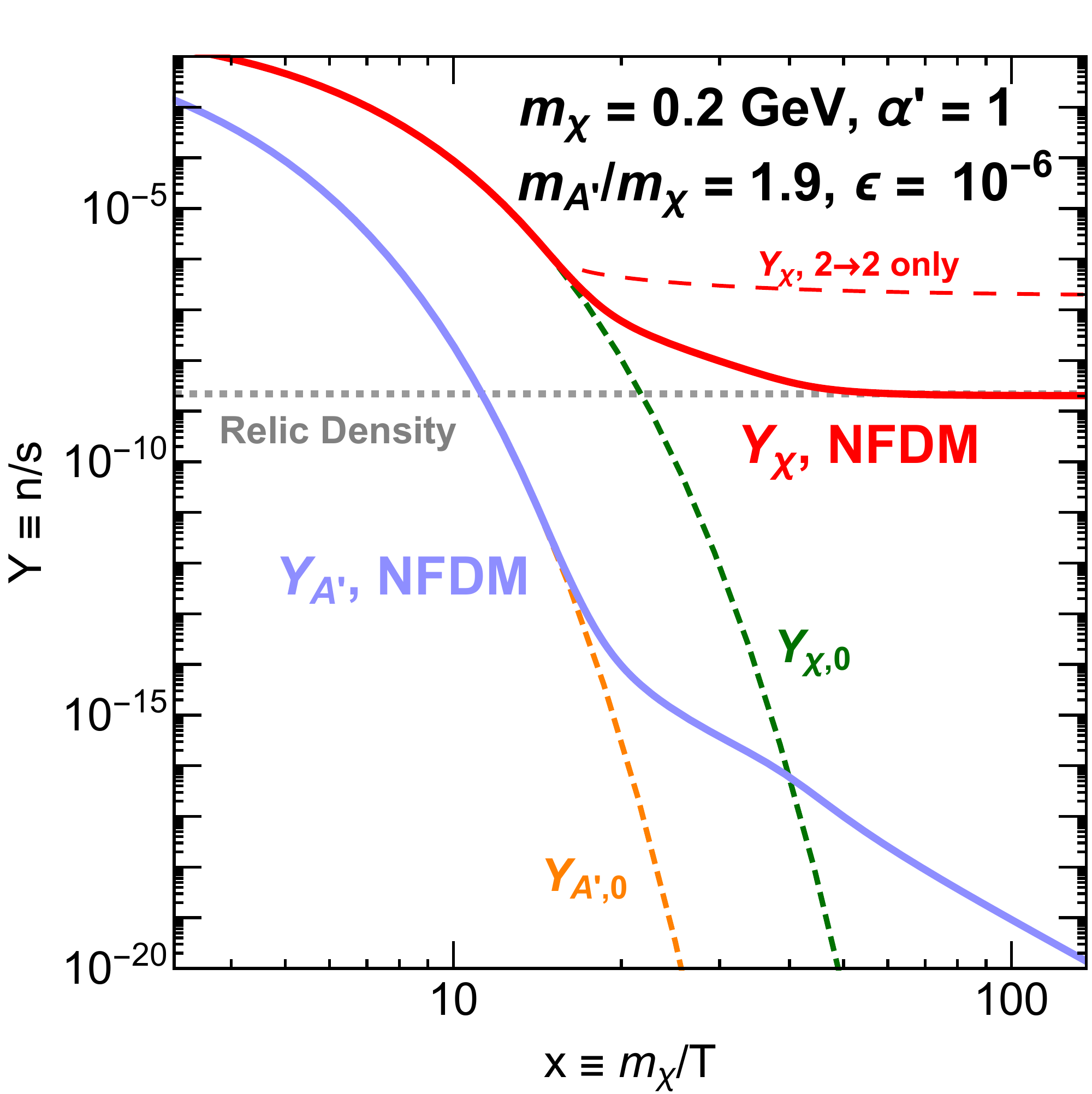}
}
\hfil
\subfloat[]{
\label{fig:23decaycontour}
\includegraphics[scale=0.56]{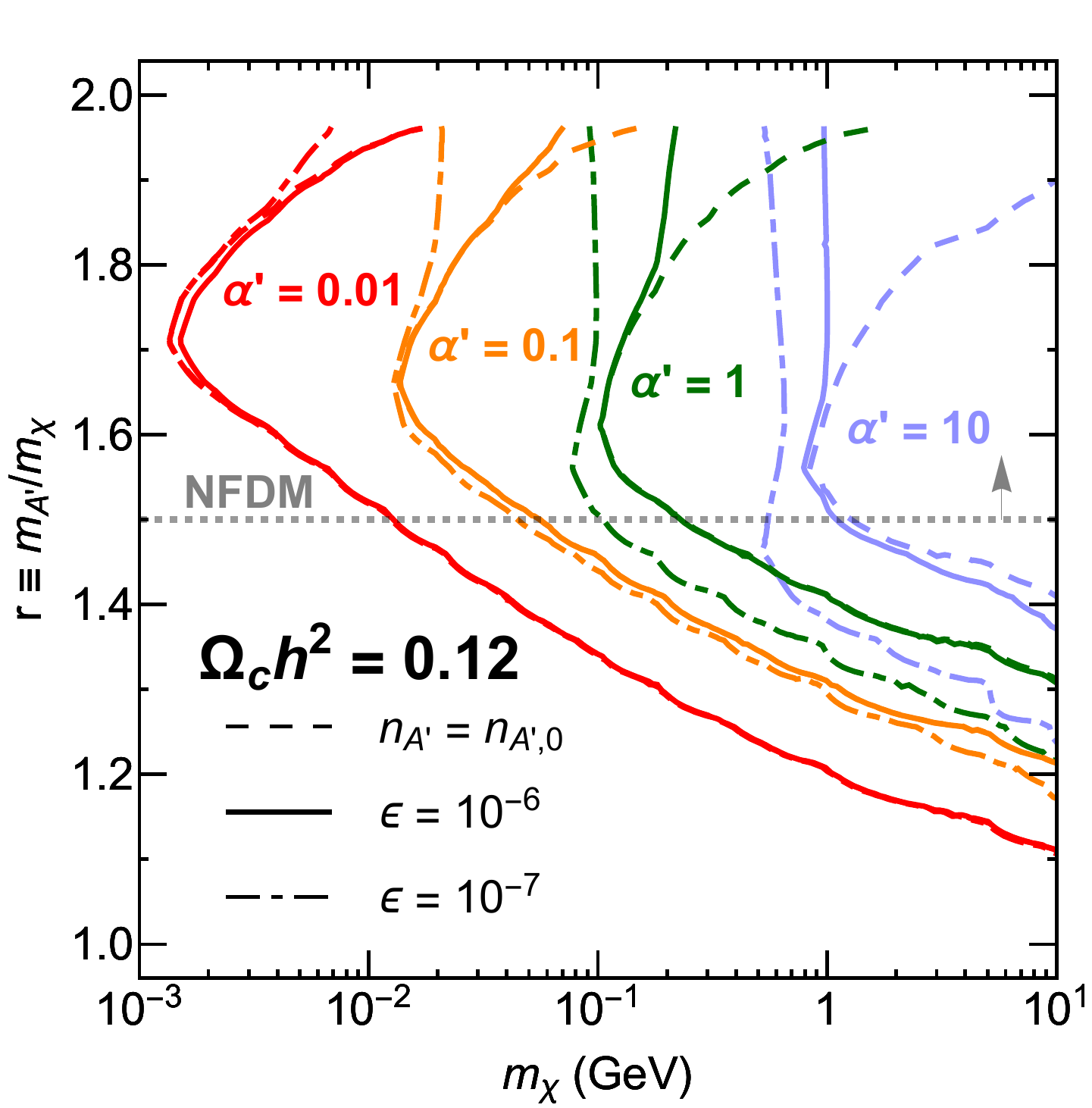}
}
\caption{Relic density in the NFDM scenario, assuming 
kinetic equilibrium of the dark sector with the SM. (a) The evolution
of the energy density of $\chi$ (red) and $A'$ (blue) for all
processes (bold) and the corresponding energy density of $\chi$
excluding the $3 \to 2$ process (red, dotted). The equilibrium
distribution of $\chi$ (green) and $A'$ (orange) are also shown for
reference;  (b) contours of the observed
present-day relic density in the $m_\chi$-$r$ plane for different
values of the coupling constant $\alpha'$.}
\label{fig:23decayall}
\end{figure*}

For sufficiently large $\epsilon$, the scattering process
$\chi e^\pm\to \chi e^\pm$  
is fast enough to keep the dark and visible sectors in kinetic 
equilibrium, $T_{d} = T_{\text{SM}}$.  By comparing the  rate inferred from the $\chi e^\pm \to \chi e^\pm$ cross section to the Hubble rate $H$ at DM freezeout, we estimate the condition to be 
\begin{equation}   
   \epsilon \gtrsim 10^{-8} \left( \frac{0.1}{ \alpha'}    \right)^{1/2}
         \left( \frac{m_\chi} { 1\, \mathrm{GeV} } \right)^{1/2} \ , 
\label{eq:thermaleq}
\end{equation}   
taking $x_f \sim 20$, the $e^\pm$ to be
relativistic,  and 
$m_{A'} \simeq m_{\chi}$.
This leaves a significant range of 
$\epsilon \lesssim 10^{-8}-10^{-4}$, depending upon $m_\chi$, such 
that
$A'$-mediated annihilations $\chi\bar\chi\to e^+e^-$ are out of equilibrium
(a requirement of our scenario), as will be shown below.

We take the dark sector masses to be in the ranges $m_\chi \lsim 10 ~
\mathrm{GeV}$ and $ m_\chi \lsim m_{A'}  < 2 m_{\chi}$.  The lower
bound on $m_{A'}$ makes $\chi \bar{\chi} \to A' A'$ kinematically
inaccessible, while the upper bound forbids the $A'\to\chi\bar\chi$ decay
channel. If $m_{A'} > 2 m_\chi$, the number-changing process
$\chi\chi\bar\chi\to A' \chi$ effectively becomes number-conserving,
$\chi\chi\bar\chi\to \chi\chi\bar\chi$.
In terms of the parameter $r\equiv m_{A^\prime}/m_\chi$, 
the relevant range is thus $1 \lesssim r \lesssim 2$.

It is enlightening to compare the equilibrium  rates (per $\chi$
particle) of the $3\to 2$ and $2\to 2$ reactions in eq.
(\ref{eq:boltz1}-\ref{eq:boltz2}),  $\Gamma_{\chi\chi\bar{\chi}\atop 
\to \chi A'} \sim \langle \sigma v^2 \rangle_{\chi\chi\bar{\chi}\atop 
\to \chi A'} n_{\chi,0}^2$ and  $\Gamma_{A'A'\atop \to \bar{\chi}
\chi}  \sim \langle \sigma v \rangle_{A'A'\atop \to \bar{\chi} \chi} 
n_{A',0}^2 /n_{\chi,0}  $. From the exponential dependences in the
equilibrium number densities, $n_{i,0} \sim \exp (-m_i/T)$, we find that
if  $m_{A'} \gtrsim \frac{3}{2}  m_\chi$, the $3 \to 2 $ reaction will
be exponentially enhanced with respect to the $2\rightarrow 2$
reaction at low temperatures.

The Boltzmann equations are solved numerically, and the results shown
in fig~\ref{fig:23decayall}. As an example, fig. 2(a)
illustrates the evolution of the $\chi$ and $A'$ abundances as a function of
$x\equiv m_\chi/T$  with $m_\chi = 0.2\,$ GeV, gauge coupling $\alpha'=1$, 
kinetic mixing $\epsilon = 10^{-6}$ and the ratio $r = 1.9$. This example has been chosen to emphasize the importance of $3 \to 2$ scatterings, but similar results are obtained for $r \gtrsim 1.5$. Here, in the case with only $2\to2$ annihilation, the DM abundance would reach its relic value at $x_f \sim 20$; in our NFDM case, in contrast, the $3\to 2$ processes and decay of the $A'$ control the freezeout, and their interplay leads to an extended freezeout continuing to $x_f\sim 60$.
If we neglect the $3 \to 2$ process the resulting abundance is overestimated by
several orders of magnitude. It is noteworthy that $Y_{A'}$
departs from the equilibrium abundance at late times, even though the
rate for $A'\to e^+e^-$ exceeds the Hubble rate, because the $3 \to 2$ or $ 2 \to 2$ processes can also strongly affect the evolution of $n_{A'}$. 

In  fig.\ \ref{fig:23decayall}(b) we plot the contours in the
$m_\chi$-$r$ plane matching the observed relic density 
\cite{Ade:2015xua}, for several values of $\alpha'$ and $\epsilon$. We consider values of $\alpha' \leq 4 \pi$, since every loop integral introduced in a Feynman diagram typically introduces an additional factor of $\alpha/4\pi$, and so perturbativity is naively maintained for this range of $\alpha'$.
$n_{A'} = n_{A',0}$ correponds to large $\epsilon$, where the rate for $A'\to e^+e^-$ 
dominates the rates for either of the two annihilation processes that generate $A'$s.
The region $r\lesssim 1.5$ corresponds to the Forbidden DM regime, and 
ref.\ \cite{D'Agnolo:2015koa} studied this regime with the assumption of $n_{A'} = n_{A',0}$: smaller values of $\epsilon$ show increasing deviation from the relic density contours obtained from this assumption, even for $r < 1.5$. For the rest of the paper, we will focus on the NFDM region $1.5\lesssim r<2$, where the $3\to 2$ process leads to a strong transition in the behavior of the relic density contour, with the exact value of $r$ for the transition depending on the coupling constant $\alpha'$. 

Normally the DM relic density is set by the strongest annihilation 
channel, which is also the last to freeze out, 
since only a single Boltzmann equation for DM is considered.  
This applies when $\epsilon$ is large, forcing 
$n_{A'} \simeq n_{A',0}$ (dashed contours). 
These contours turn to the right as
$r \to 2$ because the $3 \to 2 $ cross section
diverges, $\langle \sigma v^2 \rangle_{\chi\chi\bar{\chi} \atop\to \chi A'  }
\propto {\alpha'}^3 m_\chi^{-5} ( r - 2 )^{-7/2} $, and
$Y_\chi \sim x_f^2 / [  m_{pl} m_\chi^2  \langle \sigma v^2
\rangle_{\chi\chi\bar{\chi} \atop \to \chi A'}]^{1/2}$. Thus obtaining the correct relic density as $r \rightarrow 2$ requires a larger value of $m_\chi$.

In contrast, for moderate values of $\epsilon$, the NFDM mechanism
applies,  where the two coupled Boltzmann equations must be 
solved together. In general, we find that typically the {\it two strongest processes} (either annihilations or decays) keep the coupled system in equilibrium until the rate for one process (per $\chi$ particle) becomes comparable to the Hubble rate, and thus any weaker processes are not relevant for determining the relic abundance. In this regime, typically the decay of $A'\to e^+ e^-$ and either the $3\to2$ or $2\to2$ annihilation are the relevant processes. 
In particular, for $r\gtrsim 1.5-1.8$, the $3\to 2$ scatterings are faster than $2\to 2$, and so they dominate the freezeout, as shown in fig.~\ref{fig:23decayall}(a). The combination of $3\to 2$ scatterings and $A'$ decays can lead to a non-equilibrium density for the $A'$ particles during the freezeout of the $3\to 2$ process if $\epsilon$ is sufficiently small (e.g. $\epsilon \sim 10^{-6}-10^{-7}$), resulting in a lengthy freezeout and an $\epsilon$-dependent relic density. This behavior corresponds
to the divergence
of the dashed and solid contours in fig.~\ref{fig:23decayall}(b)
at large $r$.

\begin{figure*}[t!]
\subfloat[]{
\label{fig:eps0}
\includegraphics[scale=0.55]{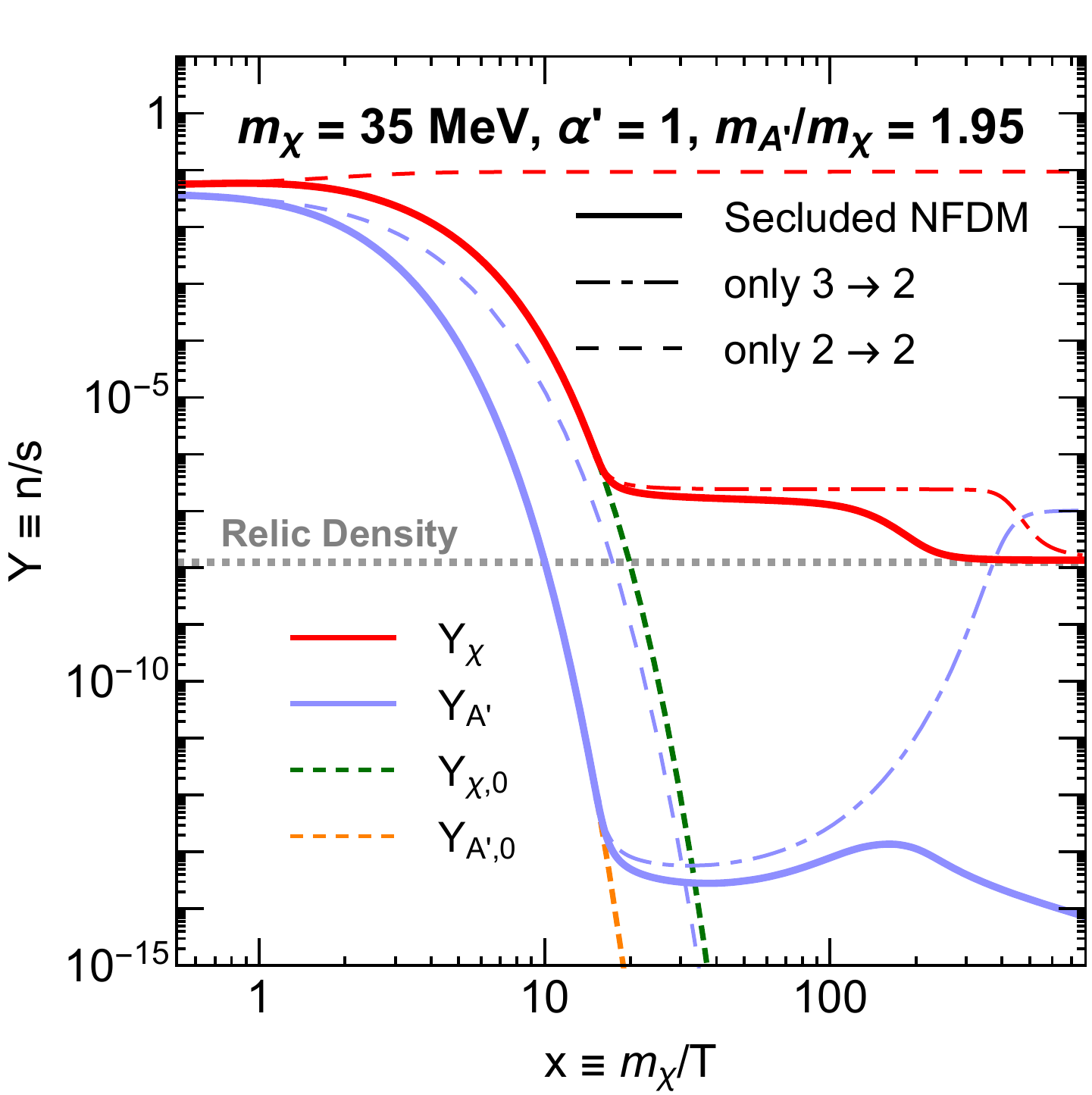}
}
\hfil
\subfloat[]{
\label{fig:eps0contour}
\includegraphics[scale=0.38]{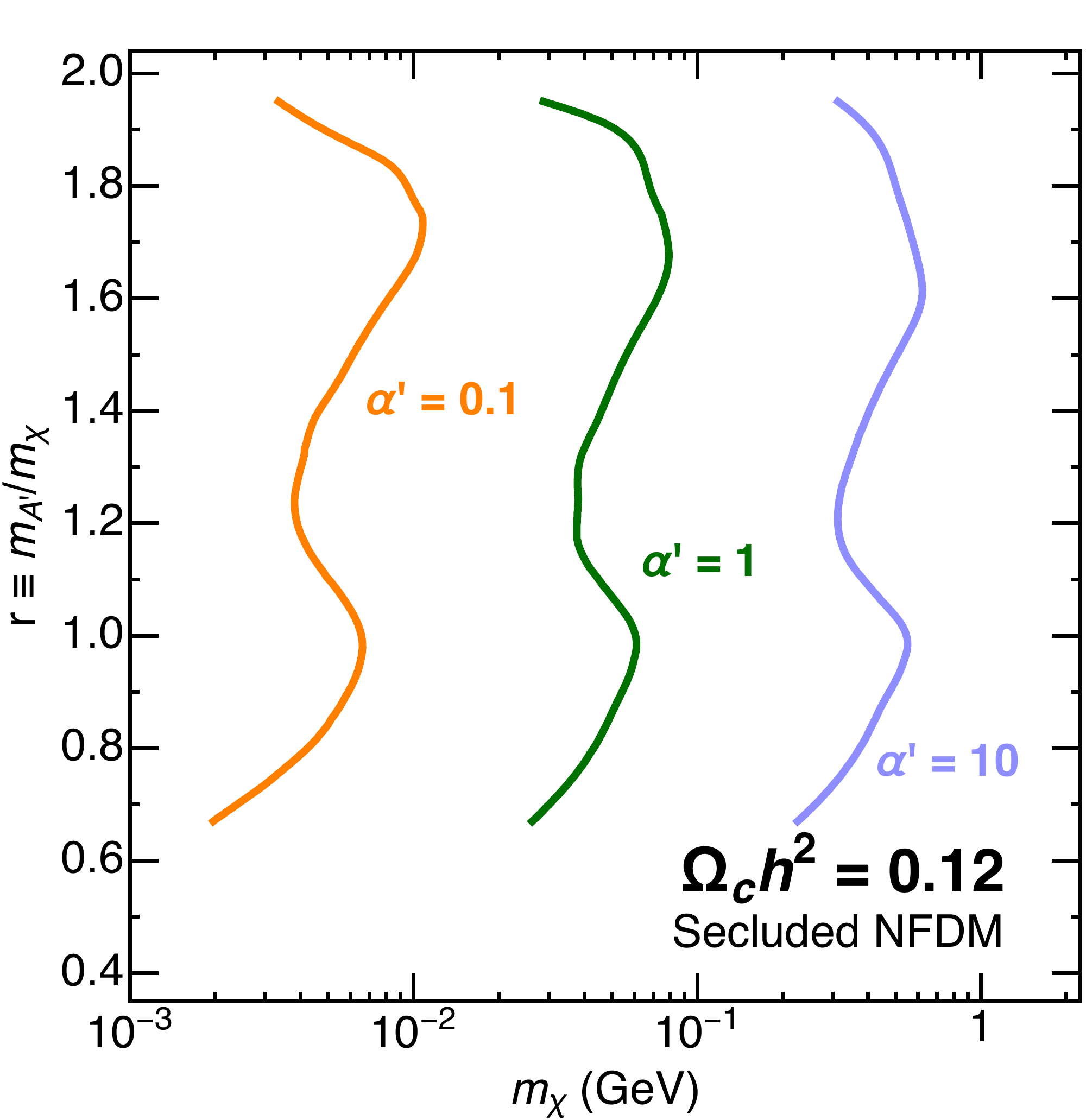}
}
\caption{NFDM, secluded hidden sector: (a) The evolution of energy
density of $\chi$ (red) and $A'$ (blue) with (solid) all relevant
processes; (dot-dashed) only $3 \to 2$ processes; (dashed) only $2 \to 2$
processes. The equilibrium distribution of $\chi$ (green) and $A'$
(orange) are also shown for reference; (b) contours of the observed
present-day relic density in the $m_\chi$-$r$ parameter space for
different coupling constants  $\alpha'$.}
\label{fig:eps0all}
\end{figure*}

%%%%%%%%%%%%%%%%%%%%%%%%%%%%%%%%%%%
\subsection{Secluded hidden sector}
\label{sec:shd}
%%%%%%%%%%%%%%%%%%%%%%%%%%%%%%%%%%%

Next we consider the limit of $\epsilon\to 0$, so that the dark 
photon is effectively stable, and the hidden sector is secluded.  
This analysis can be easily applied to multi-component DM models. Even though secluded hidden sectors are in general difficult to probe due to the lack of any interaction with the SM, they are not entirely impossible to study. Secluded hidden sectors can, for example, be constrained by the number of relativistic degrees of freedom during Big Bang Nucleosynthesis (BBN). Furthermore, in the U(1) theory considered here, the relic abundance is set by the coupling strength $\alpha'$, which in turn determines the self-interaction cross section in the dark sector. This cross section is a prediction of the model, and has observable consequences for structure formation, which can in principle be highly constraining.

Moreover, the secluded case is a useful limit that gives insight into the region of parameter space where $\epsilon$ is small but non-zero, so that kinetic equilibrium cannot be maintained with the SM. Despite the small couplings to the SM, this regime can still be effectively probed by observations of the cooling of SN1987a ~\cite{Dent:2012mx}. The secluded limit is also highly instructive as an illustration of the rich behavior that can occur in the $U(1)$ vector portal DM model when the $2 \to 2$ and $3 \to 2$ annihilations are the dominant processes at freezeout. 

To avoid warm or hot dark matter \cite{Pappadopulo:2016pkp}, we assume that $\chi$ couples additionally to some relativistic degree of freedom $\phi$ until freezeout, strongly enough to maintain
thermal equilibrium in the dark sector so that the DM temperature
redshifts with the Hubble expansion in the conventional manner,
$T\sim 1/a$.  However, the coupling of $\phi$ to $\chi$ should be
sufficiently weak that annihilation of $\chi\bar\chi\to \phi\phi$ is
negligible during freezeout, to make the NFDM freezeout mechanism dominate over
conventional $2\to 2$ annihilation. 

For a concrete model of how this can be achieved, we take $\phi$ to be a light scalar charged under some additional $U(1)$ symmetry, interacting with the dark sector through the dimension-5 operator $(1/\Lambda) \overline{\chi} \chi \phi^* \phi$. The $T \sim 1/a$ dependence is maintained by $\chi \phi \to \chi \phi$ scatterings, which has a rate that scales as $n_\phi \langle \sigma v \rangle_{\chi\phi \to \chi \phi}$, while the $\chi \overline{\chi} \to \phi^* \phi$ rate scales as $n_\chi \langle \sigma v \rangle_{\chi \overline{\chi} \to \phi^* \phi}$. To obtain a parametric estimate for a value of $\Lambda$ that would maintain both $T \sim 1/a$ and subdominance to the $2 \to 2$ and $3 \to 2$ processes considered in eq.~(\ref{eq:boltz1}) and (\ref{eq:boltz2}), we take $\langle \sigma v \rangle_{\chi\phi \to \chi \phi} \sim \langle \sigma v \rangle_{\chi \overline{\chi} \to \phi^* \phi} \sim 1/\Lambda^2$, and look for values of $\Lambda$ which ensure that the $\chi \overline{\chi} \to \phi^* \phi$ annihilation rate is subdominant up to the point of freeze-out of the two main processes. This condition is most difficult to satisfy in the case where $r = 2$, and the $2 \to 2$ rate becomes highly suppressed. Nevertheless, we find that in this limit, a suitable range of $\Lambda$ is $m_\chi^{4/3} m_{\text{pl}}^{2/3} \lesssim \Lambda^2 \lesssim m_\chi m_{\text{pl}}$, which for GeV dark matter corresponds to $10^6 \lesssim \Lambda/\text{GeV} \lesssim 10^9$. 

More generally, the dark sector has its own temperature $T_d$ which need not be the
same as that of the visible sector, $T_{\text{SM}}$; it is determined by
details of the thermal cosmological history such as the efficiency of
reheating into the dark sector after inflation.  The relic abundance
in this case depends upon the unknown parameter $\gamma \equiv T_d/T_{\text{SM}}$, 
but in a simple way: $Y_\chi \propto \gamma^{p(r)}$
where $p(r)\sim 1.6-1.8$ depends upon the mass ratio 
$r = m_{A'} / m_{\chi}$.  Here we illustrate the case of $\gamma=1$.

The evolution of $n_\chi$ and $n_{A'}$ for the secluded dark sector
is shown in 
fig.\ \ref{fig:eps0all}(a), taking $m_\chi =
35\,$MeV, $\alpha'=1$, $r=1.95$ as an example to illustrate the important interplay between the $2 \to 2$ and $3 \to 2$ interactions. 
Keeping only the $3\to 2$ reaction would predict that 
$A'$ becomes the dominant DM component, whereas in 
reality it remains highly subdominant.  
Again the freezeout process is prolonged, 
starting with the decoupling of $2\to 2$
scatterings at $x\sim 20$, while the $3\to 2$ reactions decouple
at $x\sim 150$.  Interestingly, $n_{A'}$ temporarily grows
between these two times, allowing the $2\to 2$ rate to
come back above $H$ just before freezeout completes.

In fig.~\ref{fig:eps0all}(b), we plot contours corresponding to
the observed thermal relic density in the $m_\chi$-$r$
plane for different values of $\alpha'$.  In the following we 
give a brief explanation of the contour shapes in the regions
$r \lesssim 1$, $1 \lesssim r \lesssim 1.5$ and  $1.5\lesssim r\lesssim 2$, which each show a distinct qualitative behavior:

(1) $ r \lesssim 1$. Being lighter than $\chi$, $A'$ is the dominant DM
constituent. The fastest process in this mass range is the $2 \to 2$ process $\chi \overline{\chi} \to A' A'$. Significantly below $r = 1$, the second fastest process is $3A' \to \chi \overline{\chi}$, since $n_{A',0} > n_{\chi,0}$. Near the threshold, with $n_{A',0} \sim n_{\chi,0}$, all of the other possible $3 \to 2$ processes ($\chi \chi A' \to \chi \chi$, $\chi \overline{\chi} A' \to \chi \overline{\chi}$, $\chi \overline{\chi} A' \to A' A'$, $\chi A' A' \to \chi A'$, as well as $\chi \chi \overline{\chi} \to \chi A'$ plus any conjugate processes) become important. The relic abundance curves in fig. \ref{fig:eps0all} are computed with all of these processes taken into account in the complete Boltzmann equations shown in eq. (\ref{eq:fullboltz1}) and (\ref{eq:fullboltz2}).

(2) $1 \lsim r \lsim 1.5$. $\chi$ is the dominant DM 
component.  The fastest reaction is $A'A'\leftrightarrow\chi\bar\chi$,
and it enforces $n_{A'} = n_{A',0}n_\chi/n_{\chi,0}$ during the
freezeout, and the second fastest reaction is now
$\chi\chi\bar{\chi}  \to \chi A'$, which
determines the DM abundance.   The $3\to 2$ rate goes as $n_\chi^2
\langle\sigma v^2\rangle$, which depends only weakly on $r$ through
the phase space. Therefore there is no strong correlation between the
abundance and $r$ in this region. 

(3) $1.5\lsim r \lsim 2$. $\chi$ is the dominant DM  component, but
now its abundance is determined by the two freezeout events
$A'A'\to\chi\bar\chi$ (whose rate becomes comparable to Hubble at later times) followed by  $\chi\chi\bar{\chi}  \to \chi A'$.  At large
$r\lesssim 2$,  just before freezeout completes, both reactions are
faster than $H$, allowing one to estimate the freezeout times. Taking
the $2 \to 2 $ and $ 3 \to 2 $ rates $\sim H$, and  $n_{A'} \simeq
n_{A',0}\,n_\chi/n_{\chi,0}$ enforced by fast $3 \to 2$ scatterings, we
can analytically derive contours consistent with the numerical results.

\begin{figure}[t]
\centering
\includegraphics[scale=0.55]{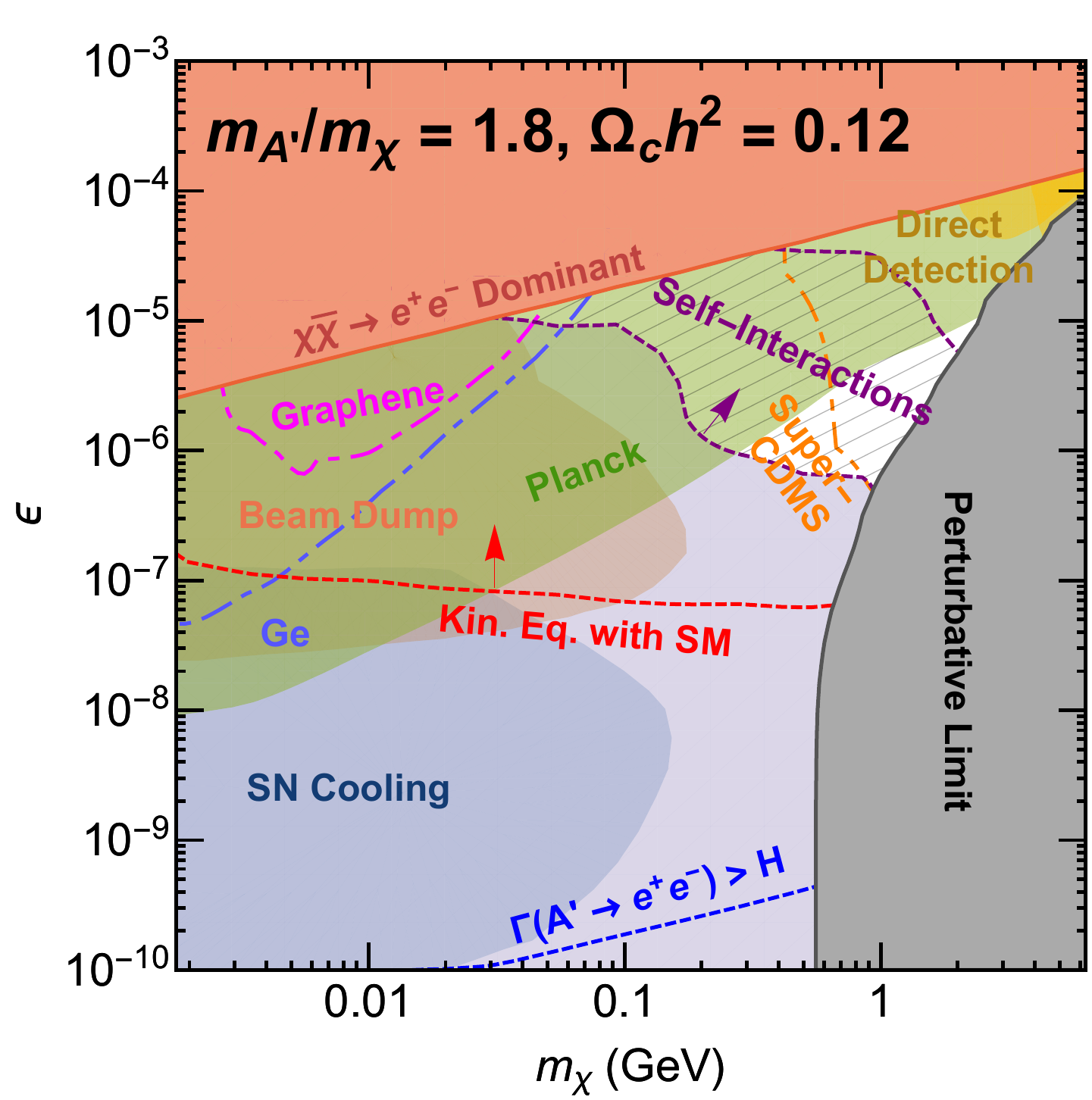}
\caption{Constraints in the $m_\chi$-$\epsilon$ plane for the case
of $m_{A'}/m_\chi = 1.8$, with  $\alpha'$ chosen to produce the
observed relic density. The allowed region is shown in white. The upper-left shaded region (red) indicates where freezeout
is dominated by the conventional $\chi\bar\chi\to e^+e^-$
annihilations.
Limits are derived from the
CMB power  spectrum \cite{Ade:2015xua} (green), beam dump 
experiments~\cite{Bjorken:2009mm,Andreas:2012mt} (pale orange),
SN1987a  cooling ~\cite{Dent:2012mx} (blue), direct 
detection~\cite{Tan:2016zwf,Akerib:2015rjg,Agnese:2015nto} (yellow) and
perturbativity, $\alpha'\geq 4\pi$ (gray). The projected
reach of SuperCDMS ~\cite{Agnese:2016cpb} (orange
dot-dashed line), electron ionization of graphene ~\cite{Hochberg:2016ntt} (magenta dot-dashed line) and germanium in a low-threshold experiment ~\cite{Lee:2015qva} (blue dot-dashed line) are also shown. The curve near $\epsilon\sim 10^{-7}$ indicates where kinetic equilibrium  with
SM is established (red dashed line). The region of parameter space where the self-interaction cross section exceeds current limits ($\sigma/m_\chi > 1 \; {\rm cm^2/g}$) (purple), and the region where the self-interaction cross section can potentially solve the small-scale structure problems 
($0.1 \; {\rm cm^2/g} < \sigma/m_\chi < 1 \; {\rm cm^2/g}$) (purple dashed lines) are displayed. The purple arrow points into the region allowed by self-interaction bounds, above and to the right of the line. The $A'$ decay rate is
faster than $H$ at freezeout above the lowest (blue) curve.
}
\label{fig:constraints}
\end{figure}

%%%%%%%%%%%%%%%%%%%%%%%%%%%%%%%%%%%
\section{Constraints}
\label{constraints}
%%%%%%%%%%%%%%%%%%%%%%%%%%%%%%%%%%%

The parameter space of NFDM is constrained by a variety of
experimental observations: i) dark photon limits coming from the cooling of
SN1987a~\cite{Dent:2012mx}; ii) similar bounds from beam dump
experiments~\cite{Bjorken:2009mm,Andreas:2012mt}; iii) limits on the
thermally-averaged cross section of $\chi \bar{\chi} \to e^+e^-$
deduced from the CMB power spectrum measured by
Planck~\cite{Ade:2015xua,Finkbeiner:2011dx,Slatyer:2009yq,Slatyer:2015jla,Liu:2016cnk},
and iv) direct detection constraints on the dark matter-nucleon
scattering cross section from PandaX-II~\cite{Tan:2016zwf}, LUX~\cite{Akerib:2015rjg} and
CDMSLite~\cite{Agnese:2015nto}. Although we have only assumed a coupling to electrons in much of this analysis for simplicity, these direct detection limits are relevant to the vector-portal DM model considered here.

Future direct detection experiments including SuperCDMS SNOLAB ~\cite{Agnese:2016cpb}, as well as electron scattering off germanium ~\cite{Derenzo:2016fse,Essig:2011nj,Lee:2015qva,Essig:2015cda} and graphene ~\cite{Hochberg:2016ntt} are also shown in the same plot.  Other current limits from XENON10 ~\cite{Essig:2017kqs}, indirect detection~\cite{Essig:2013goa}
and lower bounds on $m_\chi$ from $N_\text{eff}$~\cite{Boehm:2013jpa}
are sub-dominant to the current constraints presented here and are not shown.

Fig.~\ref{fig:constraints} summarizes these constraints  in the
$m_\chi$-$\epsilon$ plane for the illustrative value of $r
= 1.8$, with $\alpha'$ fixed to give the correct present-day relic
density, subject to the perturbativity constraint $\alpha' \leq 4\pi$.
At a large $\epsilon$ and small $m_\chi$, conventional freezeout from 
$\chi \bar{\chi} \to e^+ e^-$ annihilations dominates over the NFDM
mechanism, but this is ruled out by the CMB constraint. The
approximately horizontal red dashed contour shows the minimum value of
$\epsilon$ for which the visible and dark sectors are in kinetic
equilibrium, estimated in eq.~(\ref{eq:thermaleq}).

Self-interactions between dark matter particles with a cross section
$\sigma_\text{SI}\sim 0.1 \lesssim \sigma_\text{SI}/m_\chi \lesssim 1
\text{ cm}^2 \text{ g}^{-1}$ can potentially resolve the core-cusp and
the too-big-to-fail problems of small-structure formation with 
cold DM ~\cite{Rocha:2012jg, Spergel:1999mh,
Zavala:2012us} while remaining consistent with experimental
constraints, which set an upper bound of between 1 - 2 $\text{ cm}^2 \text{ g}^{-1}$ ~\cite{Harvey:2015hha,Robertson:2016xjh,Tulin:2017ara}. A DM mass of $m_\chi \sim (0.1 -
1)$\,GeV with $\epsilon \sim 10^{-7} - 10^{-6}$ in our model leads to a
velocity-independent self-interaction cross section that lies within this range, and can
provide a possible solution to both puzzles (though recent analysis of
clusters indicates some preference for a velocity-dependent cross
section \cite{Kaplinghat:2015aga}).  The preferred region is
between the purple dashed lines in fig. \ref{fig:constraints}, while
the cosmologically constrained region is shown in purple.

%%%%%%%%%%%%%%%%%%%%%%%%%%%%%%%%%%%
\section{Summary and Outlook} 
\label{sec:Summary}
%%%%%%%%%%%%%%%%%%%%%%%%%%%%%%%%%%%

We have demonstrated a novel scenario called {\it Not-Forbidden Dark Matter}, where an allowed  $3 \to 2 $ 
annihilation process compensates for its conventional $2\to 2$
counterpart being kinematically forbidden during thermal freezeout.
This mechanism can be potentially important in a variety of hidden
sector models, including vector-portal, scalar-portal and composite
DM. The DM mass and the mediator (or second DM)
mass are of the same order, which would naturally arise in a hidden
sector characterized by a single scale.  

Taking the vector-portal DM model as an example,  we found that
in some parts of the NFDM  parameter space, the combined effect of $ 3
\to 2 $, $ 2 \to 2$ and $A'$ decay channels is to significantly prolong the period
of freezeout.  The commonly-neglected $3 \to 2$ annihilation channel
can change the predicted relic density by orders of magnitude. 
Although this model is restricted by an abundance of experimental
constraints, viable examples remain in the mass range $\sim
(0.1-1)\,\mathrm{GeV}$, with a self-interaction cross section that is
coincidentally of the right order for solving the small scale
structure problems of  $\Lambda$CDM cosmological simulations. This is
a well-motivated target for future direct detection~\cite{Agnese:2016cpb,Essig:2011nj,Derenzo:2016fse} 
and dark photon searches~\cite{Ilten:2015hya,Ilten:2016tkc,Alekhin:2015byh,Gardner:2015wea,Moreno:2013mja}.

%%%%%%%%%%%%%%%%%%%%%%%%%%%%%%%%%%%
%%%%%%%%%%%%%%%%%%%%%%%%%%%%%%%%%%%
\bigskip

While we were completing this work, \cite{Dey:2016qgf} appeared, presenting a related idea.
Their work focuses on keV-MeV scalar DM and requires additional scalar
``assister" particles.

\section*{Acknowledgements}
We thank Yonit Hochberg, Lina Necib, Nicholas Rodd, Joshua Ruderman and Yotam Soreq for useful discussions. We also thank A.D. Dolgov for translating his early work on DM $3 \to 2$ annihilations \cite{dolgov1980concentration,Dolgov:2017ujf} and bringing it to our attention.
JC is supported by NSERC (Canada) and FRQNT (Qu\'ebec).
HL, TS and WX are supported by the U.S.
Department of Energy under grant Contract Numbers 
DE-SC-00012567 and DE-SC-0013999.
%%%%%%%%%%%%%%%%%%%%%%%%%%%%%%%%%%%
%%%%%%%%%%%%%%%%%%%%%%%%%%%%%%%%%%%

\appendix
%\section*{Supplemental Material}
%Here we provide further details concerning the distinctive 
%features of the NFDM freezeout mechanism, and how it differs from
%the usual $2\to 2$ annihilation scenario.

\section{Coupled Boltzmann equations and prolonged freezeout}

As mentioned above, an essential difference between NFDM and
conventional DM freezeout is the importance of tracking the evolution
of both the DM $\chi$ and the mediator particle (in our model, $A'$),
by solving the coupled Boltzmann equations (eq. (\ref{eq:boltz1}) and (\ref{eq:boltz2}) for relevant terms when $r \gtrsim 1$, eq. (\ref{eq:fullboltz1}) and (\ref{eq:fullboltz2}) for the complete equations)  for their respective densities.   The presence of two
equations implies that more than one scattering (or decay) process can
be important for determining the final abundance; hence both the
fastest and second fastest reactions are typically relevant.

This is in contrast to conventional DM freezeout based upon a single
Boltzmann equation, where the abundance depends upon the strongest
channel.  In the large $\epsilon$ limit of our model,
$A'$ decay is the fastest process, and enforces equilibrium of 
$A'$, $n_{A'} = n_{A',0}$. 
Hence smaller values of $\epsilon$ are necessary to realize the rich cosmology that comes from the interplay of the coupled Boltzmann equations of $\chi$ and $A'$. To simplify the subsequent discussion, we assume that these $\epsilon$-suppressed reactions are negligibly slow, i.e. we work in the secluded dark sector regime of the NFDM model.

It is useful to define the 
net rate of $3 \lra 2$ or $2 \lra 2$ interactions per $\chi$ or
$A'$ particle by considering the collision terms in the Boltzmann equations, 
written in the form $R_{\chi} \equiv d \log n_{\chi}/dt = -3H - R_{\chi}(3 \leftrightarrow 2) + R_{\chi}(2 \leftrightarrow 2)$ and $R_{A'} \equiv d \log n_{A'}/dt = -3H + R_{A'}(3 \leftrightarrow 2) - R_{A'}(2 \leftrightarrow 2)$, where
\begin{eqnarray}
\label{eq:Rnchi1}  
    R_{\chi} ( 3 \leftrightarrow 2 )  
         &\equiv&   2 \frac{n_{A'} }{ n_\chi} R_{A'} ( 3 \leftrightarrow 2 )   \\
      &=&
         \frac{1}{4} \langle \sigma v^2 \rangle_{\chi\chi\bar{\chi}  \to \chi A'  }
      \left(  n_\chi^2 - n_{\chi,0}^2 \frac{n_{A'}} {n_{A',0}}\right)\nonumber    \\
	&\equiv& R_\chi(\chi\chi\bar{\chi}  \to \chi A') - R_\chi(\chi A'\to\chi\chi\bar{\chi} ) \nonumber,
     \end{eqnarray}
\begin{eqnarray}
      \label{eq:Rnchi2}
    R_{\chi} ( 2  \leftrightarrow 2 )  &\equiv&
             \frac{n_{A'} }{ n_\chi} R_{A'} ( 2 \leftrightarrow 2 )  \\
      &=&
      \langle \sigma v \rangle_{A'A' \to \bar{\chi} \chi}
      \left(  \frac{n_{A'}^2}{n_\chi}  -  \frac{ n_{A', 0}^2  n_\chi}{n_{\chi,0}^2}\right)\nonumber\\
	&\equiv&  R_\chi(A'A' \to \bar{\chi} \chi) - R_\chi(\bar{\chi} \chi \to A'A') \nonumber.
\end{eqnarray}
Likewise, we define $2(n_{A'}/n_\chi)R_{A'}(\chi \chi \bar{\chi} \to \chi A') \equiv R_\chi(\chi \chi \bar{\chi} \to \chi A')$ and so on for the unidirectional rates. In this way, the signs for these definitions have been chosen so that all of the rates of individual sub-processes are now positive, although the overall rates $R_\chi$ and $R_{A'}$ can have any sign. When $m_{A'} > m_\chi$ and $T < m_\chi,m_{A'}$, the lower density of $A'$ relative to $\chi$
implies that $R_\chi (  3 \leftrightarrow 2) $ ($R_\chi (  2 \leftrightarrow 2) $) is generally smaller in magnitude than 
$R_{A'} (  3 \leftrightarrow 2) $ ($R_{A'} (  2 \leftrightarrow 2) $).
Thus the rates $R_{\chi}$ tend to fall below $H$ earlier than the corresponding rates $R_{A'}$.
This separation between freezeout of $\chi$ and $A'$ is the origin of the prolonged duration of the
overall freezeout process.

Suppose that only one channel, for example $2 \to 2$, occurs fast enough such that $R_\chi(A'A' \to \bar{\chi} \chi)\gg H$;
then this rate tends to be nearly equal to that of the reverse reaction, $R_\chi(\bar{\chi} \chi \to A'A')$, 
enforcing the condition $ n_{A'}^2  \simeq  n_{A', 0}^2\, {n_\chi^2}/{n_{\chi,0}^2}$
(though the cancellation is imperfect, so that the total rate $R_\chi(2\lra 2)$ is also typically greater than $H$).
This by itself is not sufficient to force both the $\chi$ and $A'$ densities to track their
equilibrium values. For that, one generically needs
both $R_\chi(3\leftrightarrow 2)>H$ and $R_\chi(2\leftrightarrow 2)>H$ so that both independent combinations 
$n_\chi-n_{\chi,0}$ and $n_{A'}-n_{A',0}$
are driven to zero.\footnote{The typical behavior is that the strongest process is such that both the forward
and backward rates exceed $H$, as well as their difference.  For the second-strongest, only one of these
need be greater than $H$.}
This is always true at  sufficiently early times, allowing us to use equilibrium initial conditions for the
Boltzmann equations. The DM density $n_\chi$ starts to deviate from equilibrium when the rate of the weaker
annihilation channel becomes comparable to $H$; hence the second-strongest channel initiates the
freezeout process.

\begin{figure*}[t!]
\subfloat[]{
\label{fig:rate1n}
\includegraphics[scale=0.51]{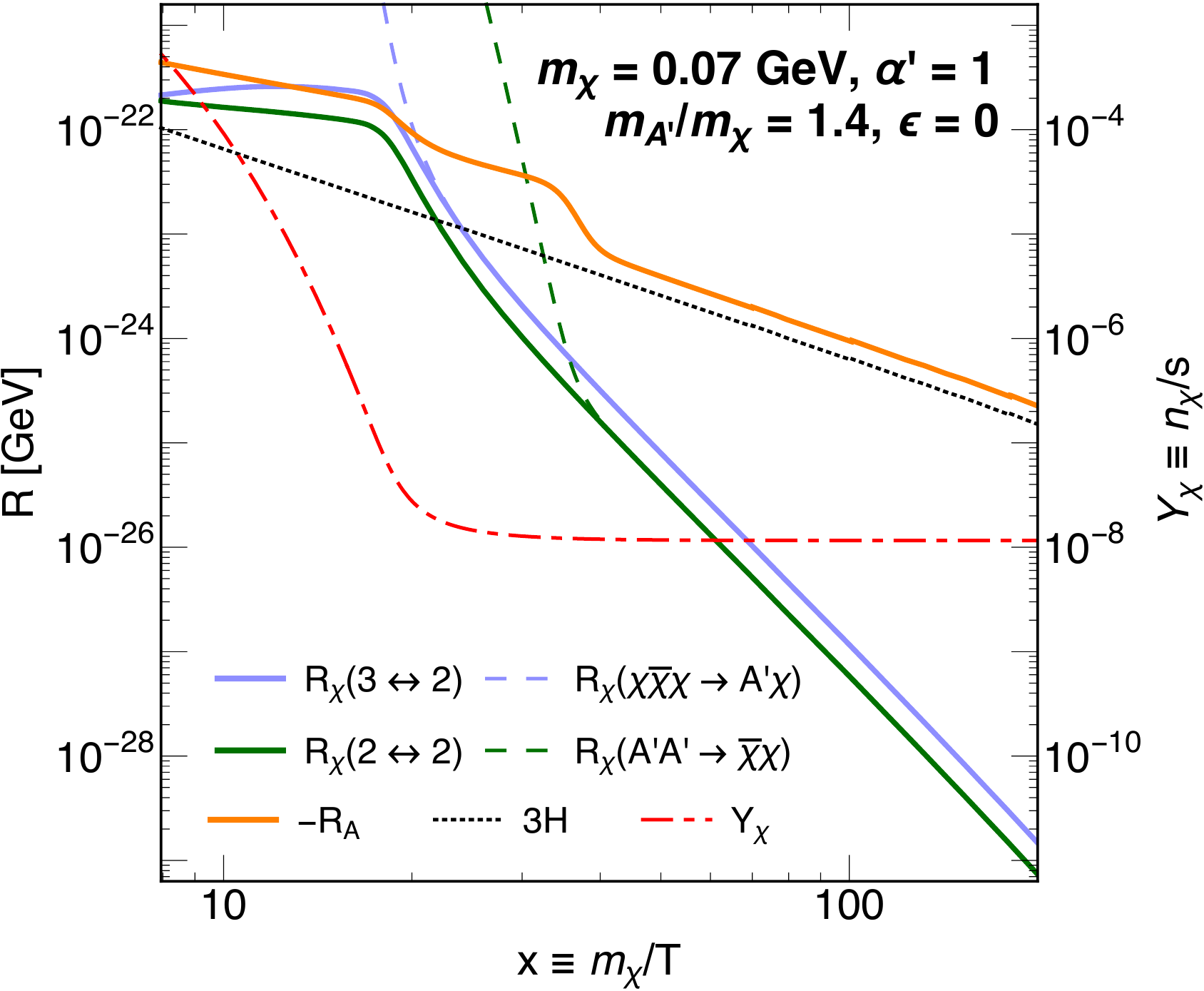}
}
\hfil
\subfloat[]{
\label{fig:rate1A}
\includegraphics[scale=0.52]{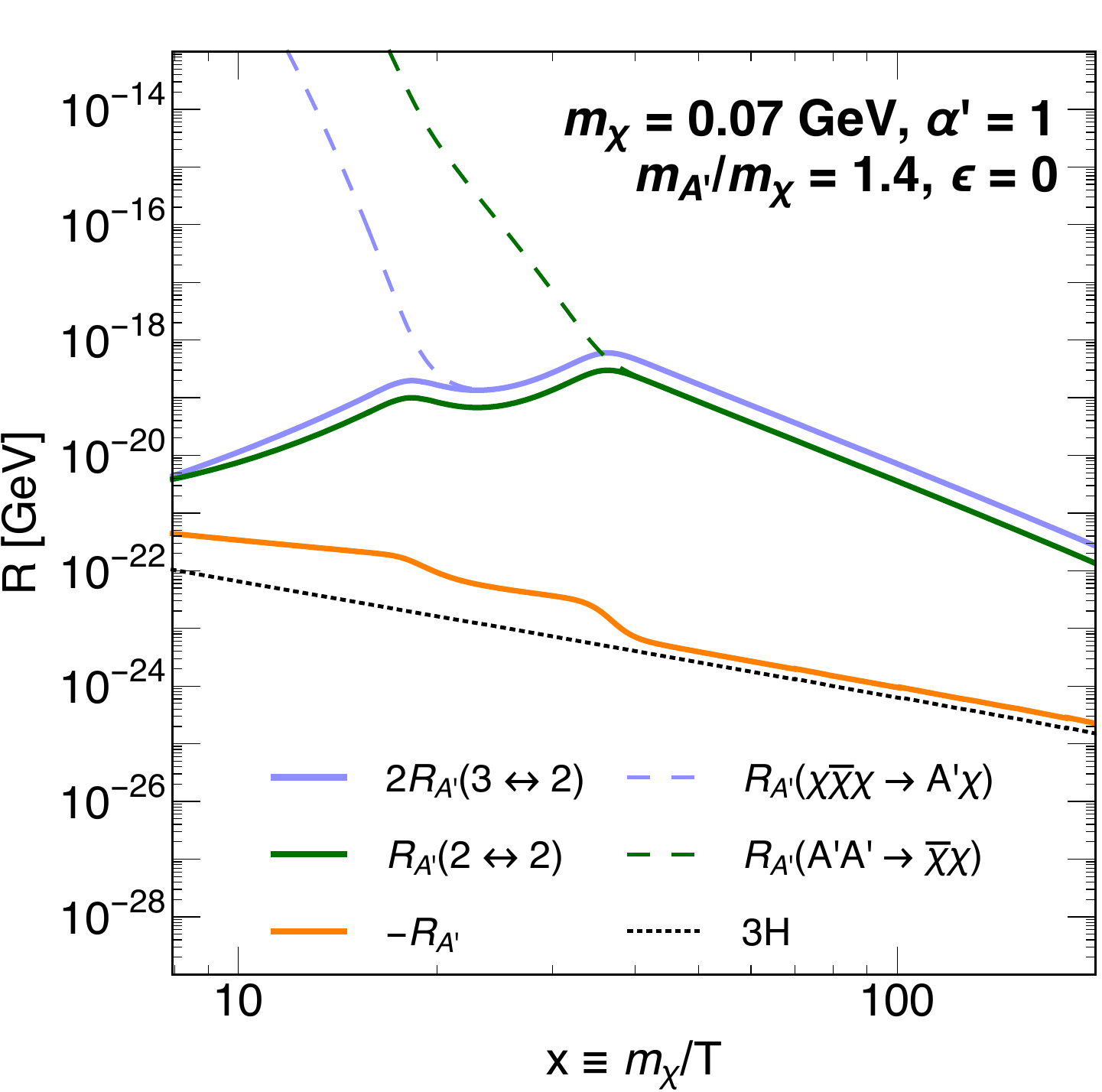}
}
\caption{Rates of different processes during freezeout for $m_{A'}/m_\chi = 1.4$: (a) evolution of $R_\chi(3
\leftrightarrow 2)$ (light blue), $R_\chi(2 \leftrightarrow 2)$ (green) and the $A'$ total rate $R_{A'}$
(orange) as a function of $x \equiv m_\chi / T$. Rates for processes in one direction, $R_\chi(\chi \chi
\bar{\chi} \to A' \chi)$ (light blue, dashed) and $R_\chi(A' A' \to \chi \bar{\chi})$ (green, dashed) are also
shown. The dark matter abundance $Y_\chi$ (red, dot-dashed) is plotted, with the appropriate (dimensionless) units
given on the right-hand axis; (b) evolution of $R_{A'}(3 \leftrightarrow 2)$ (light blue), $R_{A'}(2
\leftrightarrow 2)$ (green) and the $A'$ total rate $R_{A'}$ (orange) as a function of $x$. Rates for processes
in one direction, $R_{A'}(\chi \chi \bar{\chi} \to A' \chi)$ (light blue, dashed) and $R_{A'}(A' A' \to \chi
\bar{\chi})$ (green, dashed) are also shown. The evolution of the Hubble rate $H$ (black, dotted) is
shown in both plots for reference.}
\label{fig:rate1all}
\end{figure*}

\begin{figure*}[t!]
\subfloat[]{
\includegraphics[scale=0.5]{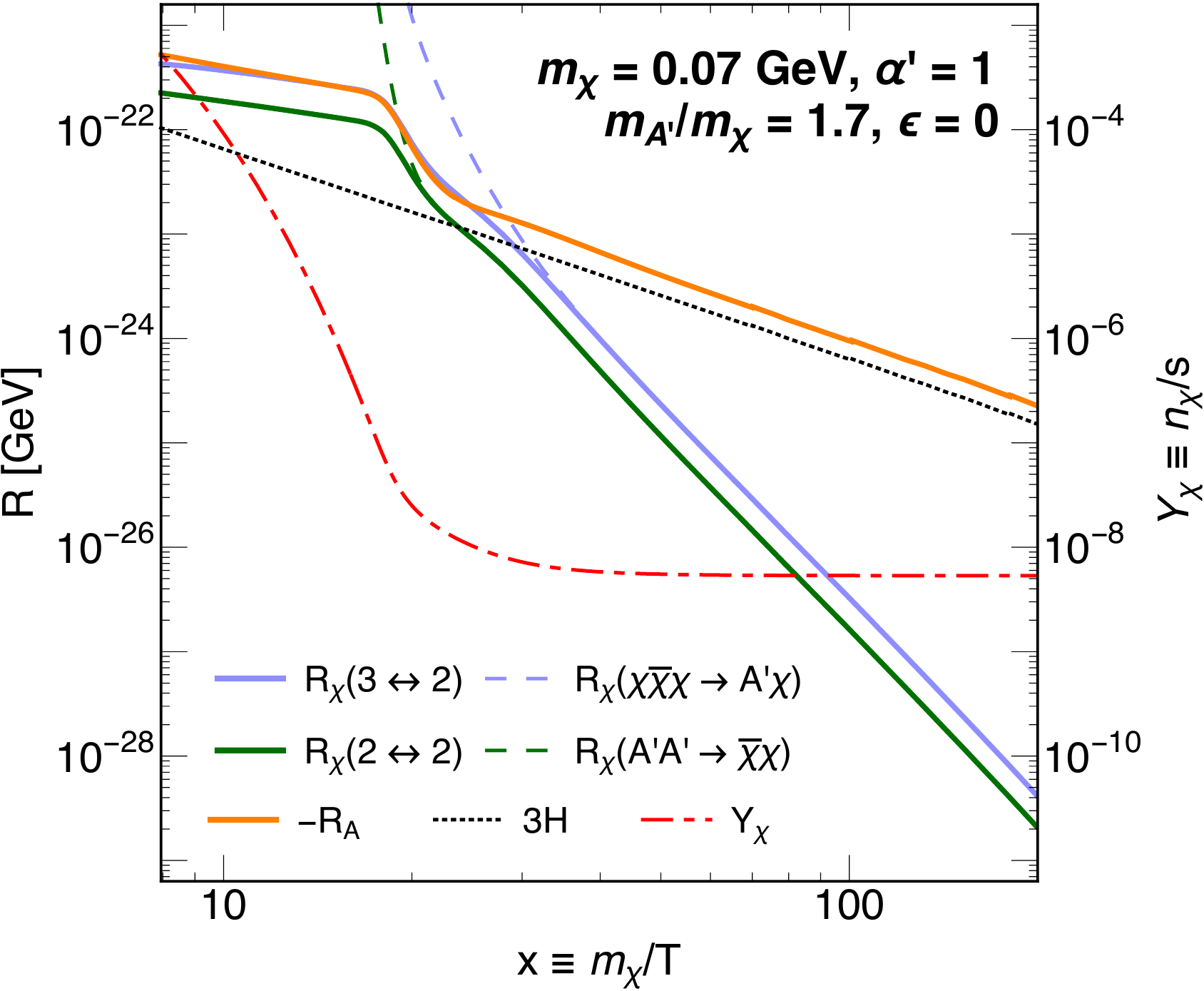}
} 
 \hfil
\subfloat[]{
  \includegraphics[scale=0.5]{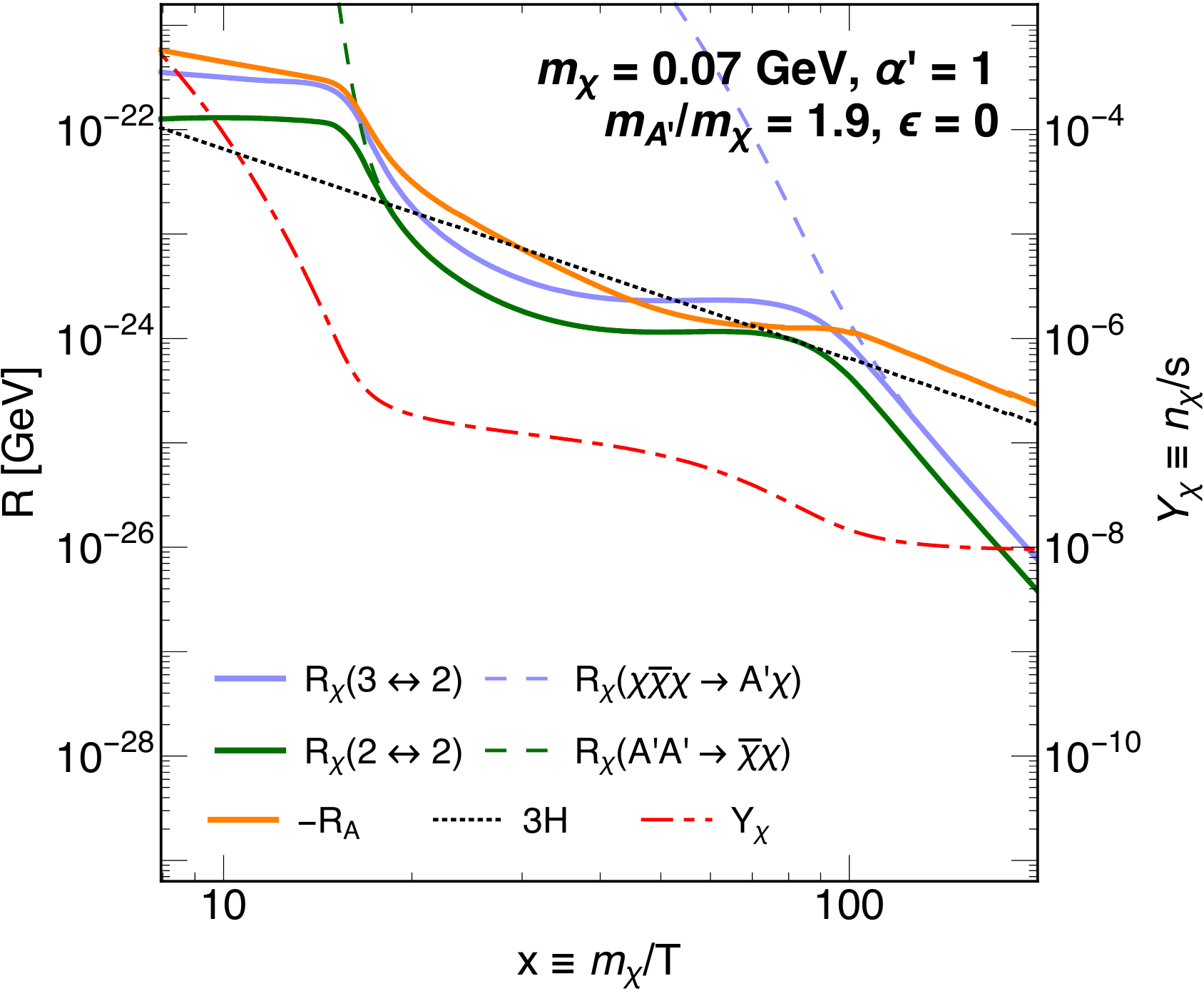}
}
\caption{Same as fig. \ref{fig:rate1n}, but with (a) $m_{A'}/m_\chi = 1.7$, and (b) $m_{A'}/m_\chi = 1.9$.}
\label{fig:rate2n}
\end{figure*}

To illustrate this behavior,  we show some examples of the evolution of the rates in
fig.~\ref{fig:rate1n}, \ref{fig:rate1A} and \ref{fig:rate2n}. Each example has  the same DM mass $m_\chi =
70~\mathrm{MeV}$, coupling $\alpha' =1 $, and kinetic mixing $\epsilon = 0$, but  different  values of $ r
= 1.4$, $ 1.7$, $ 1.9$. In these figures, the dot-dashed lines corresponding to  the evolution of DM number
density are shown to highlight the time of DM freezeout.  For $ r = 1.4$, the freezeout period is relatively
short; for $ r= 1.7$, freezeout is prolonged; and  for $ r = 1.9$, the freezeout is prolonged further and may
indeed be thought of as two separated  freezeouts. 

In fig.~\ref{fig:rate1A}, we show the two rates $R_{A'} ( 3
\leftrightarrow  2 )$ and $R_{A'} ( 2 \leftrightarrow  2 ) $, which are
much larger than $H$; these cancel each other to order $H$. The
behavior is similar for other values of $r$. Since  $R_{A'} ( 3
\leftrightarrow  2 )   \simeq  R_A ( 2 \leftrightarrow  2 ) $, 
eq.~(\ref{eq:Rnchi1}, \ref{eq:Rnchi2}) implies that $  R_\chi (
3 \leftrightarrow  2 ) \simeq 2 R_\chi ( 2 \leftrightarrow  2 ) $. This
relation is demonstrated in fig.~\ref{fig:rate1n} and ~\ref{fig:rate2n}.

Comparing these net rates however does not tell us which process
controls freezeout. Instead, we should look at the dashed lines,
which indicate  the unidirectional rates, $R_\chi ( A'A' \to
\bar{\chi} \chi )$  and  $R_\chi( \chi\chi\bar{\chi}  \to \chi A' )$.
Processes are out of equilibrium when these dashed lines overlap with
the solid lines. From the unidirectional rates, we can identify the
weaker annihilation channel and thus which process initiates the
freezeout. For $r = 1.4$, the weaker process is $3 \to 2$, and 
fig.~\ref{fig:rate1n} confirms that the freezeout is indeed triggered
by $ 3 \to 2 $. For $ r = 1.7$ and $ r = 1.9$, the dashed line for 
$R_\chi ( A'A' \to \bar{\chi} \chi )$ in fig.\ \ref{fig:rate2n} merges
with the solid line, $R_\chi ( 2 \leftrightarrow  2 ) $, when the rate
is  about $3 H$. It is the weaker channel $ 2 \to 2$ that initiates
freezeout. 

One difference between $ r < 1.5 $ and $ r > 1.5$ in
figs.~\ref{fig:rate1n} versus \ref{fig:rate2n} is that  $ r > 1.5$ normally
has a longer freezeout. The duration depends upon
whether the rate of the weaker annihilation channel is sensitive to
$n_{A'}$. For $r > 1.5$, the weaker process $ 2 \to 2$ has the rate $R_\chi (A' A' \to \chi \chi)
\sim  \langle \sigma v  \rangle_{A'A' \to \bar{\chi} \chi} n_{A'}^2 /
n_\chi $. Prior to the final freezeout, the larger $ 3 \to 2 $ rate
imposes the constraint that $n_{A'} \simeq n_{A',0} n_\chi^2 /
n_{\chi,0}^2 \sim r^{3/2} x^{3/2} m^{-3}   \exp( (2- r )x ) n_\chi^2$.
Since $n_{A'}$ increases with time, this means that the $ 2 \to 2 $
rate  $R_\chi ( A'A' \to \bar{\chi} \chi )$ can  be kept at the same
order as $H$ for a long period.  For this reason, the freezeout is
prolonged. For $ r < 1.5$, the duration  is relatively
short, because the rate of the weaker $ 3 \to 2$ process  goes as  $R_\chi(
\chi\chi\bar{\chi}  \to \chi A' ) \propto n_\chi^2$,  where $n_\chi$
is decreasing with time.

Armed with our insight that the second-strongest channel matters
critically for freezeout, and having understood the reason that the
freezeout process is longer for $ r \gtrsim 1.5$, we can also explain the shape of the contours in
fig. 2(b) in the main text. As discussed briefly in the main text, there are several important regimes:

\begin{enumerate}[leftmargin=2\parindent]
\item For $n_{A'} = n_{A',0}$, corresponding to large $\epsilon$, the two Boltzmann  equations are
reduced to one, and the shape of the contours can be understood using the usual parametrics of DM freezeout. This behavior occurs for the contours overlapping the dashed contours in fig. 2(b).

\item For $\epsilon=0$, we recover the secluded case discussed above, where the interplay of the $3\leftrightarrow 2$ and $2\leftrightarrow 2$ processes controls the freezeout. This behavior also occurs in the region where $r \sim 1.5$ and $ \alpha'  =10$,
because the $3 \to 2$ and $2 \to 2 $ rates for DM are significantly larger than $ R_\chi( A' \to e^+ e^-) \equiv \Gamma ( A' \to e^+ e^- ) n_{A'} / n_{\chi}$, so that  $\Gamma_{A' \to e^+ e^-}$ can be neglected.\footnote{We define the $A'$ decay rate $ R_{\chi} ( A' \to e^+ e^-)$ with
respect to the DM density; even though
this quantity does not appear in Boltzmann equation of DM, the
coupling to the $A'$ Boltzmann equation will cause $A' \to e^+ e^-$
to play an important role in determining the rate for DM processes in
some cases.}

\item For moderate $\epsilon$, the rates of the three processes should be
compared in order to determine which is weakest, and hence
irrelevant to the DM freezeout. The relevant rates to compare
are $R_\chi(\chi \bar{\chi} \chi \to A' \chi )$, $R_{\chi}( A' A' \to \chi \bar{\chi} )$ and 
$ R_{\chi} ( A' \to e^+ e^-)$,
evaluated at the Hubble crossing time of the
annihilation processes. Consider the case where $ 2 \to 2$ has a
lower rate than $3 \to 2$, such that it falls below $H$
first.  Whether the DM density freezes out or not at this time depends on
the relative sizes of $ R_{\chi} ( A' \to e^+ e^-)$    and $R_{\chi}( 
A' A' \to \chi \bar{\chi} )$. When $R_{\chi}( A' A' \to \chi \bar{\chi} )  <  R_{\chi} ( A' \to e^+ e^-) $, the
larger $R_{\chi} (A' \to e^+ e^-)$ rate in the coupled Boltzmann
equations provides enough constraints to keep $n_\chi$ and $n_{A'}$ 
near their equilibrium values.
An example of this more complex case is shown in
fig. 2(a) of the main text, where the
freezeout starts when  $R_\chi( 3 \leftrightarrow 2 ) \sim H$.
Using the Boltzmann equation of $A'$, this rate is determined by the 
$A'$ decay, $R_\chi( 3 \leftrightarrow 2 ) \sim  2 R_\chi( A' \to e^+
e^-) $;  freezeout then terminates when $R_\chi( \chi
\bar{\chi} \chi \to A' \chi ) \sim H$. 

More broadly, this case is realized when $ \epsilon =
10^{-6}$, and either $r$ is close to $2$, or $\alpha'$ is large and $ r > 1.5 $. Fig. 2(b) shows that in this region the $\epsilon = 10^{-6}$ contours (solid lines) do not overlap with the dashed contours, 
for which the constraint $n_{A'} = n_{A',0}$ is imposed. In this regime the $ 3 \to 2 $ rate is the largest, and when 
$R_\chi( A' A' \to \chi \bar{\chi}) \sim H$,  $ R_{\chi} ( A' \to e^+ e^-)>R_\chi( A' A' \to \chi \bar{\chi} )$.
The freezeout is thus controlled by $ 3 \to 2$ processes and the decay of $A'$. In this case the $A'$ decay rate is not fast enough to keep the $A'$ abundance in equilibrium, and both $n_{A'}$ and $n_\chi$ are increased during freezeout relative to their values when the $A'$s remain in equilibrium. The $\chi$ annihilation rate thus needs to be increased to maintain the correct relic density, requiring lower $\chi$ masses; this is the reason that the contours in fig. 2(b) bend toward lower masses as $\epsilon$ is decreased, for large $r$.

\end{enumerate}

\section{Dependence on $T_d$} 

If the dark sector is secluded, its temperature $T_d$ may differ from
that of the visible sector, $T_{\text{SM}}$.  This difference
affects the evolution of the $\chi$ and $A'$ densities and ultimately
the DM relic abundance. In the Boltzmann equations
(eq. (2, 3) in the main text), taking $T_d \neq T_\text{SM}$
changes the equilibrium densities, so that $n_{\chi,0} \sim \exp(-
m_\chi/T_d) = \exp(- x/\gamma)$, where we have defined $\gamma \equiv
T_d/T_\text{SM}$, and $x$ is still given by $x \equiv
m_\chi/T_\text{SM}$. Likewise, $n_{A',0} \sim \exp(-r x/\gamma)$.
Keeping in mind that $H$ is determined by $T_\text{SM}$, we can solve
the Boltzmann equations for $n_\chi(x)$ and $n_{A'}(x)$ with the 
$\gamma$-dependence coming from the equilibrium
densities. 

\begin{figure}
\includegraphics[scale=0.53]{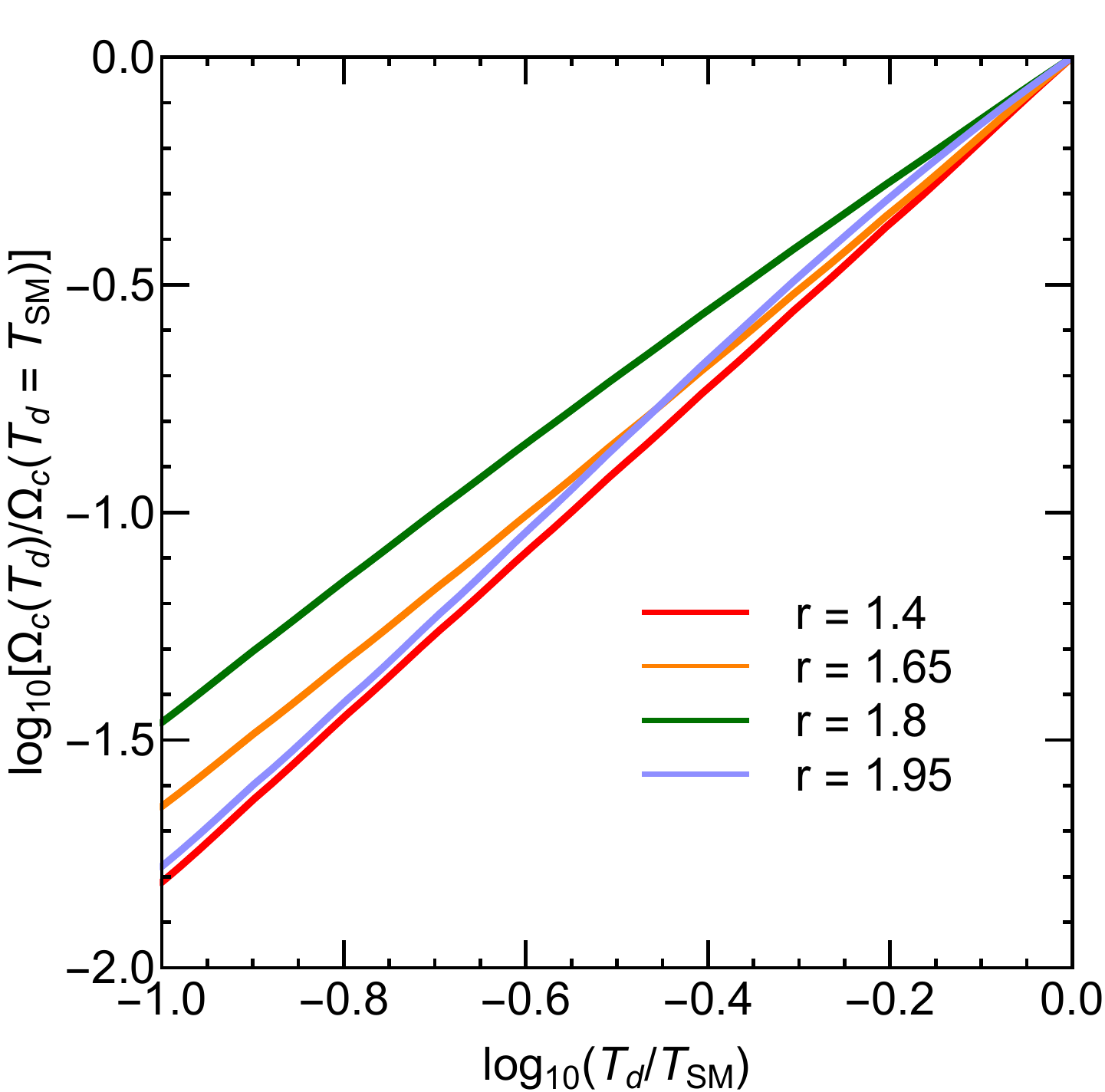}
\caption{Ratio of the relic abundance when $T_d < T_\text{SM}$ to the relic abundance with $T_d = T_\text{SM}$ as a function of $\gamma \equiv T_d/T_\text{SM}$ for $r \equiv m_{A'}/m_\chi = 1.4$ (red), 1.65 (orange), 1.8 (green) and 1.95 (light blue).}
\label{fig:omegaRatio_vs_Td}
\end{figure}

Fig.\ \ref{fig:omegaRatio_vs_Td} shows the behavior of the ratio of
relic abundances $\Omega_c(T_d)/\Omega_c(T_d = T_\text{SM})$ as a
function of $T_d$ for $0.1\le\gamma\le 1$. Having $T_d < T_\text{SM}$
leads to an earlier freezeout, since the exponential decrease in the
equilibrium densities occurs more rapidly. For values of $r$ where the
backward and forward $3 \to 2$ processes fall out of equilibrium at
freezeout, we expect that $n_{\chi}^2 \sim H/ \langle \sigma v^2
\rangle_{\chi \chi \bar{\chi} \to \chi A'} \sim 1/x_f^2$, and
therefore that the relic abundance scales as $\Omega_c \sim x_f^2$.
For $1< r \lesssim 1.5$ where the $3 \to 2$ process determines the DM
abundance, the exponential dependence of $n_{\chi,0}$ with $x/\gamma$
results in $\Omega_c \sim \gamma^2$. On the other hand, in the case of
$r \lesssim 2$, the second freezeout occurs at $n_\chi \propto
n_{\chi,0}^4/n_{A',0}^2$, and a similar argument leads again to
$\Omega_c \sim \gamma^2$. At intermediate values of $r$, both the $3
\to 2$ and $2 \to 2$ processes freeze out at similar times. For a $2
\to 2$ freezeout, $n_\chi \sim H/\langle \sigma v \rangle_{\bar{\chi}
\chi \to A' A'}$, and as a result $\Omega_c \sim \gamma$.
Qualitatively, we expect the $\gamma$ dependence to lie between these
two regimes for intermediate values of $r$.    

\section{Constraints at different $r$}

\begin{figure*}[!]
\subfloat[]{
  \label{fig:constraints_r_14}
\includegraphics[scale=0.57]{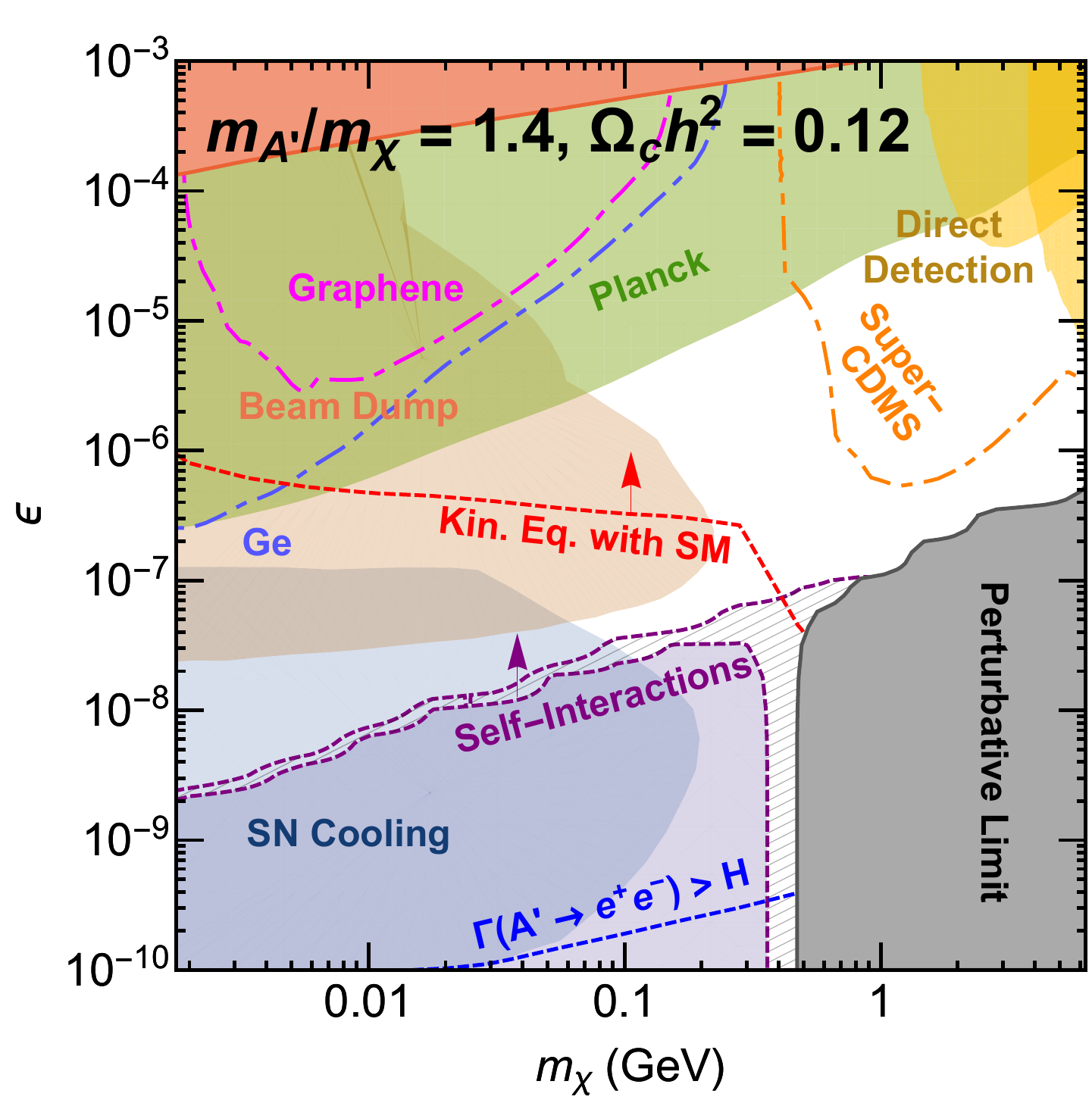}
} 
\subfloat[]{
  \label{fig:constraints_r_16}
  \includegraphics[scale=0.57]{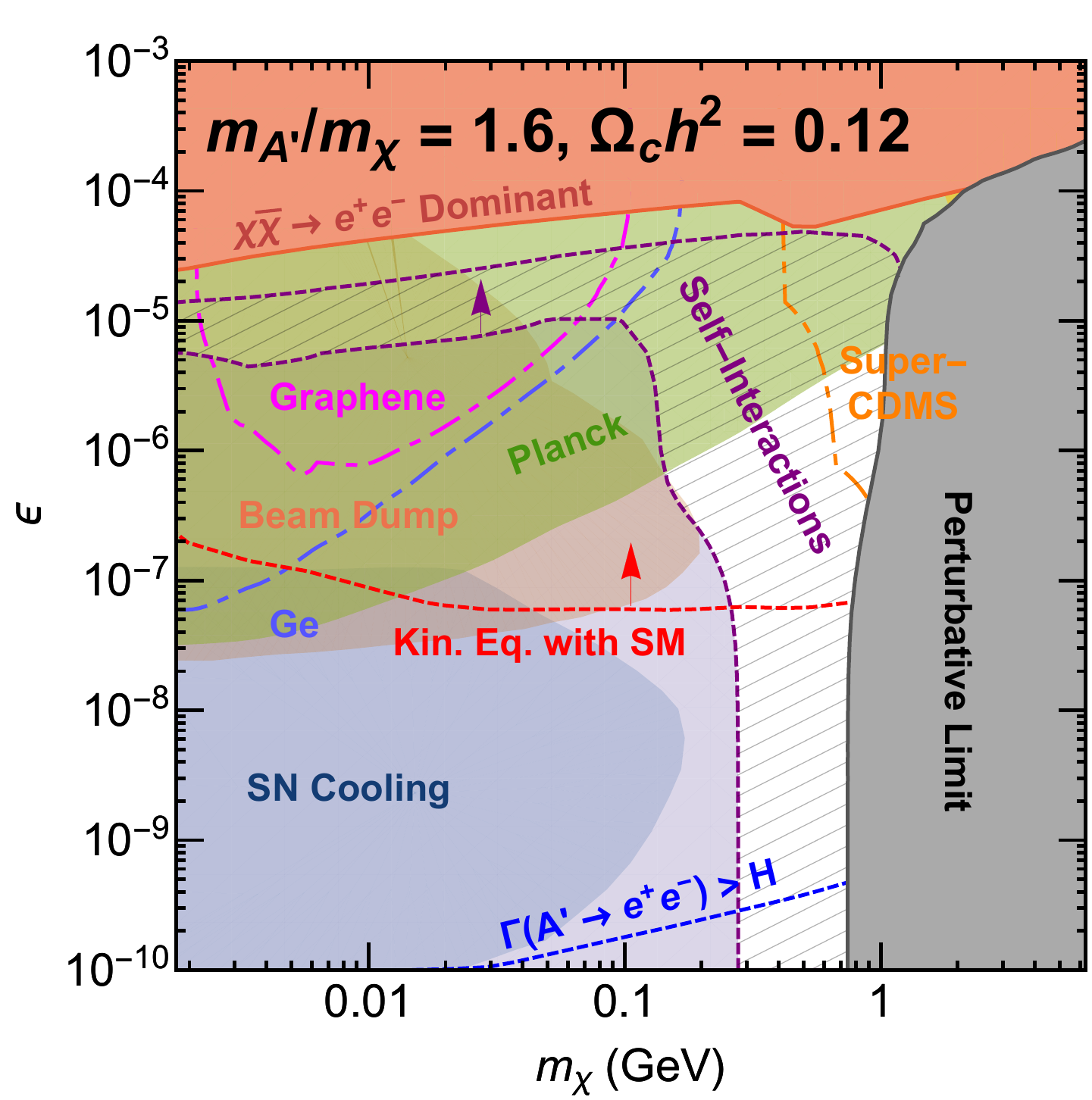}
}
\caption{Same as fig. 4 in the main text, but with (a) $m_{A'}/m_\chi = 1.4$, and (b) $m_{A'}/m_\chi = 1.6$.}
\label{fig:constraintsSupp}
\end{figure*}

Fig.~\ref{fig:constraintsSupp} shows the constraints in the $m_\chi-\epsilon$ plane for
two representative values of $r$, with $\alpha'$ fixed to give the
correct present-day relic density. These have the same general
features as in fig. 4 in the main text, but also exhibit several
distinct characteristics that we explain here. 

At $r = 1.4$, the transition from the secluded limit ($\epsilon \to
0$) to the kinetic equilibrium limit occurs in the range
$\epsilon \sim 10^{-9} - 10^{-6}$, which leads to a rapid decrease in $\alpha'$ between these two phases at a fixed value of $m_\chi$. This explains the rapid
change in the behavior of the region with suitable self-interaction for $m_\chi \lesssim 100\text{ MeV}$.

At $r = 1.6$, the most distinctive feature is the change in behavior
of the region where annihilations to $e^+e^-$ dominates at $m_\chi
\sim 100\text{ MeV}$. At masses smaller than this point, the $2 \to 2$
process is freezes out last, while at larger masses, it is the $3 \to
2$ process which does so. This difference accounts for the change in
the slope of the boundary. There is no such transition for the other
two cases, since at $r = 1.8$, the $3 \to 2$ process always freezes
out last, while for $r = 1.4$ it is the $2 \to 2$ process instead. 

For all values of $r$, a significant part of the $m_\chi - \epsilon$
parameter space is still consistent with the present-day relic density
while evading experimental constraints, showing that the NFDM scenario
is robust against taking different values of $r \gtrsim 1.5$.  
\vspace{2cm}

\section{Complete Boltzmann equations}
\label{app:boltz}

The complete Boltzmann equations, including all relevant $2 \to 2$ and $3 \to 2$ processes for the full range of $r$ considered is:
\begin{multline}
    \frac{d n_{\chi}}{dt} + 3H n_{\chi} = -\frac{1}{4} \langle \sigma v^2 \rangle_{\chi \chi \overline{\chi} \atop \to \chi A'} \left(n_\chi^3 - n_{\chi,0}^2 n_\chi \frac{n_{A'}}{n_{A',0}}\right) \\
    + \langle \sigma v \rangle_{A' A' \atop \to \overline{\chi} \chi} \left(n_{A'}^2 - n_{A',0}^2 \frac{n_\chi^2}{n_{\chi,0}^2}\right) \\
    - \frac{1}{2} \langle \sigma v^2 \rangle_{\chi \overline{\chi} A' \atop \to A' A'} \left(n^2 n_{A'} - n_{\chi,0}^2 \frac{n_{A'}^2}{n_{A',0}}\right) \\
    + \frac{1}{3} \langle \sigma v^2 \rangle_{A'A'A' \atop \to \chi \overline{\chi}} \left(n_{A'}^3 - n_{A',0}^3 \frac{n_\chi^2}{n_{\chi,0}^2}\right) \\
    - \frac{1}{2} \langle \sigma v \rangle_{\chi \overline{\chi} \to e^+ e^-} \left(n_\chi^2 - n_{\chi,0}^2 \right),
 \label{eq:fullboltz1}
\end{multline}

\begin{multline}
    \frac{d n_{A'}}{dt} + 3 H n_{A'} = \frac{1}{8} \langle \sigma v^2 \rangle_{\chi \chi \overline{\chi} \atop \to \chi A'} \left(n_\chi^3 - n_{\chi,0}^2 n_\chi \frac{n_{A'}}{n_{A',0}} \right) \\
    - \langle \sigma v \rangle_{A' A' \atop \to \overline{\chi} \chi} \left(n_{A'}^2 - n_{A',0}^2 \frac{n_\chi^2}{n_{\chi,0}^2} \right) - \Gamma_{A' \to f \overline{f}} \left(n_{A'} - n_{A',0} \right) \\
    - \frac{1}{4} \left( \langle \sigma v^2 \rangle_{\chi \overline{\chi} A' \atop \to \chi \overline{\chi}} + \langle \sigma v^2 \rangle_{\chi \chi A' \atop \to \chi \chi} \right) \left(n_\chi^2 n_{A'} - n_\chi^2 n_{A',0} \right) \\
    + \frac{1}{4} \langle \sigma v^2 \rangle_{\chi \overline{\chi} A' \atop \to A' A'} \left(n_\chi^2 n_{A'} - n_{\chi,0}^2 \frac{n_{A'}^2}{n_{A',0}} \right) \\
    - \frac{1}{2} \langle \sigma v^2 \rangle_{\chi A' A' \atop \to \chi A'} \left(n_\chi n_{A'}^2 - n_\chi n_{A'} n_{A',0} \right) \\
    - \frac{1}{2} \langle \sigma v^2 \rangle_{A' A' A' \atop \to \chi \overline{\chi}} \left(n_{A'}^3 - n_{A',0}^3 \frac{n_\chi^2}{n_{\chi,0}^2} \right).
\label{eq:fullboltz2}
\end{multline}

The symmetry factors preceding each term properly account for the number of identical particles in the initial state, the net number of particles created or destroyed in each annihilation process, as well as conjugate processes. These equations are used in all numerical calculations shown in the paper. 

\section{Cross sections and decay rates}

\label{app:xsects}
The decay rate for $A'\to e^+e^-$ is
\begin{eqnarray}
   \Gamma (A' \to e^+ e^-)  &=&  \frac{ \epsilon^2 \alpha_{em}  }{3} m_{A'} 
   \left(  1 + 2 \frac{  m_e^2 }{ m_{A'}^2 }  \right)
      \sqrt{ 1 - 4 \frac{ m_e^2}{ m_{A'}^2} } \; .\nonumber\\
\end{eqnarray}

For scattering cross sections, the thermally-averaged $2 \to 2$ cross section for the process $1 + 2 \to 3 + 4$ is given by

\begin{multline}
    \langle \sigma v \rangle_{12 \to 34} = \frac{1}{S_f} \frac{1}{n_1 n_2} \int \prod_{i=1}^4 \frac{g_i d^3 \vec{p}_i}{(2\pi)^3 2E_i} \\
    \times (2\pi)^4 \delta^4(p_1 + p_2 - p_3 - p_4) f_1 f_2 \overline{|\mathcal{M}|^2},
\end{multline}
where $g_i$ is the number of degrees of freedom and $f_i$ is the phase space distribution of species $i$. The averaged squared matrix element $\overline{|\mathcal{M}|^2}$ is averaged over both the initial and final state degrees of freedom. $S_f = \prod_i n_i!$ is a symmetry factor, where $n_i$ is the number of identical particles of species $i$ in the final state. Initial state symmetry factors are included explicitly in the Boltzmann equation, eq. (\ref{eq:boltz1}) and (\ref{eq:boltz2}). This convention may differ from other sources in the literature. 

Similarly, the thermally-averaged $3 \to 2$ cross section for the process $1 + 2 + 3 \to 4 + 5$ is
\begin{multline}
    \langle \sigma v^2 \rangle_{123 \to 45} = \frac{1}{S_f} \frac{1}{n_1 n_2 n_3} \int \prod_{i=1}^5 \frac{g_i d^3 p_i}{(2\pi)^3 2E_i} \\
    \times (2\pi)^4 \delta^4(p_1 + p_2 + p_3 - p_4 - p_5) f_1 f_2 f_3 \overline{|\mathcal{M}|^2}.
\end{multline} 

For simplicity and unless otherwise stated, we give cross sections at the kinematic threshold of the respective processes. In this limit, thermally averaged cross sections are
\begin{alignat}{1}
    \langle \sigma v \rangle_{12 \to 34} = \frac{g_3 g_4}{32 \pi S_f m_1 m_2} \lambda^{1/2} (m_1 + m_2, m_3, m_4) \overline{|\mathcal{M}|^2},
\end{alignat}
and
\begin{multline}
    \langle \sigma v^2 \rangle_{123 \to 45} = \frac{g_4 g_5}{64 \pi S_f m_1 m_2 m_3} \\
    \times  \lambda^{1/2} (m_1 + m_2 + m_3, m_4, m_5) \overline{|\mathcal{M}|^2},
\end{multline}
where $\lambda(x, y, z) \equiv (1 - (z+y)^2/x^2) (1 - (z-y)^2/x^2)$. This expression agrees with the result for the specific process of $3 \chi \to 2 \chi$ computed in \cite{Kuflik:2017iqs}.\footnote{Note that different conventions are used between this paper and \cite{Kuflik:2017iqs}.}

\renewcommand{\arraystretch}{3} 

\setlength{\tabcolsep}{15pt}

\begin{table*}
\begin{tabular}{c c c} 
\toprule

Process & $\overline{|\mathcal{M}|^2}$ & Phase Space \\
\hline
$A' A' A' \to \chi \overline{\chi}$ & $\frac{g'^6(153r^6 - 47r^4 - 60r^2 + 24)}{9 m_\chi^2 r^8}$ & $\frac{\sqrt{9r^2 - 4}}{48 \pi m_\chi^3 r^4}$  \\
$\chi A' A' \to \chi A'$ & $\frac{2g'^6(195r^8 + 1156r^7 + 4670r^6 + 9444r^5 + 12214r^4 + 11192r^3 + 6732r^2 + 2272r + 320)}{9 m_\chi^2 (r+1)^2 (r+2)^4 (2r+1)(r^2 - 2r - 2)^2}$ & $\frac{3 \sqrt{3} \sqrt{3r^2 + 8r + 4}}{32 \pi m_\chi^3 r (2r+1)^2}$ \\
$\chi \chi A' \to \chi \chi$ & $\frac{2 g'^6 r(r+4)}{3 m_\chi^2(r+1)^2 (r+2)^2}$ & $\frac{\sqrt{r(r+4)}}{32 \pi m_\chi^3 r (r+2) }$  \\
$\chi \overline{\chi} A' \to \chi \overline{\chi}$ & $\frac{g'^6 (r+4) (9r^6 + 24r^5 + 4r^4 - 40r^3 + 168r^2 - 224r + 128)}{6 m_\chi^2 r^3 (r-2)^2 (r+1)^2 (r+2)^2}$ & $\frac{\sqrt{r(r+4)}}{16 \pi m_\chi^3 r (r+2)}$  \\
$\chi \overline{\chi} A' \to A' A'$ & $\frac{16g'^6(21r^6 - 4r^5 - 17r^4 + 24r^3 + 216r^2 + 288r + 112)}{27 m_\chi^2 (r-2)^4 (r+1)^4 (r+2)^2}$ & $\frac{9 \sqrt{-3r^2 + 4r + 4}}{128 \pi m_\chi^3 r(r+2)}$ \\
$\chi \overline{\chi} \chi \to A' \chi$ & $\frac{g'^6 (r-4)(r+4)(-32r^8 + 167r^6 - 534r^4 + 668r^2 - 512)}{36 m_\chi^2 (r^2 - 4)^4 (r^2 + 2)^2}$ & $\frac{\sqrt{r^4 - 20r^2 + 64}}{96 \pi m_\chi^3}$  \\
$\chi \overline{\chi} \to A' A' $ & $\frac{16g'^4(1-r^2)}{9(r^2 - 2)^2}$ & $\frac{9 \sqrt{1 - r^2}}{64 \pi m_\chi^2}$  \\
$A' A' \to \chi \overline{\chi}$ & $\frac{32g'^4(r^4 - 1)}{9r^4}$ & $\frac{\sqrt{r^2 - 1}}{8 \pi m_\chi^2 r^3}$ \\
$\chi \overline{\chi} \to e^+e^-$ & $\frac{4 e^2 \epsilon^2 g'^2 \left(2 + m_e^2/m_\chi^2\right)}{(r^2 - 4)^2}$ & $\frac{\sqrt{1 - m_e^2/m_\chi^2}}{8 \pi m_\chi^2}$ \\
\botrule
\end{tabular}
\caption{List of initial- and final-state averaged squared matrix element $\overline{|\mathcal{M}|^2}$ of each process, as well as the phase space factor $P$ such that $\langle \sigma v \rangle$ or $\langle \sigma v^2 \rangle = P \overline{|\mathcal{M}|^2}$. All values are evaluated at the kinematic threshold. For the last two processes, we use the expression for $\chi \overline{\chi} \to A' A'$ for $r < 1$ and $A' A' \to \chi \overline{\chi}$ for $r > 1$. }
\label{tab:xsec}
\end{table*}

In Table \ref{tab:xsec}, we list all of the number changing processes that are included in the Boltzmann equations eq. (\ref{eq:fullboltz1}) and (\ref{eq:fullboltz2}), the initial- and final-state averaged squared matrix element $\overline{|\mathcal{M}|^2}$ of each process as well as the phase space factor $P$ such that $\langle \sigma v \rangle$ or $\langle \sigma v^2 \rangle = P \overline{|\mathcal{M}|^2}$. We define $r \equiv m_{A'}/m_\chi$ throughout.

Two other processes that are important to our analysis are $\chi e^\pm \to \chi e^\pm$ which maintains kinetic equilibrium between the dark sector and the SM, and dark matter-dark matter scattering. 

\begin{itemize}

    \item{$\chi e^\pm \to \chi e^\pm$}: this cross section is important in determining if the DM is in kinetic equilibrium with the SM. In the limit where $T < \mu_{e\chi}$, where $\mu_{e\chi}$ is the electron-DM reduced mass, we have

    \begin{alignat}{1}
        \langle \sigma v \rangle = \frac{2(g' \epsilon e)^2 \mu_{e\chi}^2}{ \pi m_{A'}^4} \left(\frac{2T}{\pi \mu_{e\chi}}\right)^{1/2}.
    \end{alignat}
    
    At high temperatures, it approaches the limit
    
    \begin{alignat}{1}
        \langle \sigma v \rangle \to \frac{(g' \epsilon e)^2}{4 \pi m_{A'}^2}.
    \end{alignat}
    
    To get accurate results, however, we must use the exact thermal average over the cross section for $\chi e^\pm \to \chi e^\pm$, which is given by: 
    
    \begin{multline}
        \sigma = \frac{(g' \epsilon e)^2}{8\pi}  \left[ \frac{1}{s} + \frac{2}{r^2 m_\chi^2} + \frac{8m_e^2 + r^4 m_\chi^2}{r^2 [h(m_\chi,s) + r^2 m_\chi^2 s]} \right. \\
        \left. - \frac{2(r^2 m_\chi^2 + s)}{h(m_\chi,s)} \log \left(1 + \frac{h(m_\chi,s)}{r^2 m_\chi^2 s}\right)\right],
    \end{multline}
    
    where $h(m_\chi, s) = [s - (m_\chi + m_e)^2][s - (m_\chi - m_e)^2]$. The thermal average is then given by
    
    \begin{multline}
        %\langle \sigma v \rangle = \frac{1}{8T m_\chi^2 m_e^2 K_2(m_\chi/T) K_2(m_e/T)} \\
        \langle \sigma v \rangle = \int_{M^2}^\infty \frac{ds}{\sqrt{s}} \cdot \frac{h(m_\chi ,s) K_1(\sqrt{s}/T) \sigma}{8 T m_\chi^2 m_e^2 K_2(m_\chi/T) K_2(m_e/T)} \, ,
        %\cdot \int_{M^2}^\infty \frac{ds}{\sqrt{s}} \left[(s - m_\chi^2 - m_e^2)^2 - 4m_\chi^2 m_e^2 \right] K_1(\sqrt{s}/T) \sigma \; ,
    \end{multline}
    
    where $M = m_e + m_\chi$. 
    
    \item{$\chi \chi \to \chi \chi$}: the self-interaction cross section, averaged over particle-particle and particle-antiparticle scattering, is ~\cite{{D'Agnolo:2015koa}}: 
    \begin{alignat}{1}
        \frac{\sigma_{\text{SI}}}{m_\chi} = \frac{3g'^4}{16\pi m_\chi^3} \frac{16 - 16r^2 + 5r^4}{r^4(r^2 - 4)^2} \; .
    \end{alignat}
    
\end{itemize}

\bibliography{ref}

\vspace{0.2in}

\end{document}